\documentclass[english,pra,twocolumn,notitlepage,superscriptaddress,floatfix]{revtex4-1}
\pdfoutput=1
\usepackage{amstext}
\usepackage{amssymb}
\usepackage{amsmath}
\usepackage{dsfont}
\usepackage{float}
\usepackage{siunitx}
\usepackage{bbold}
\usepackage[braket,qm]{qcircuit}
\usepackage{braket}
\usepackage{mathtools}
\usepackage[dvipsnames]{xcolor}
\usepackage{tikz}
\newcommand{\tikzcircle}[2][black,fill=black]{\tikz[baseline=-0.5ex]\draw[#1,radius=#2] (0,0.03) circle ;}
\usepackage{enumitem}

\usepackage{hyperref}
\hypersetup{colorlinks=true,allcolors=Black}

\begin{document}

\title{Hardware-efficient quantum random access memory with hybrid quantum acoustic systems}

\author{Connor T. Hann}
\affiliation{Departments of Applied Physics and Physics, Yale University, New Haven, Connecticut 06511, USA}

\author{Chang-Ling Zou}
\affiliation{Key Laboratory of Quantum Information, CAS, University of Science and Technology of China, Hefei, China.}

\author{Yaxing Zhang}
\affiliation{Departments of Applied Physics and Physics, Yale University, New Haven, Connecticut 06511, USA}

\author{Yiwen Chu}
\affiliation{Department of Physics, ETH Z{\"u}rich, 8093 Z{\"u}rich, Switzerland}

\author{Robert J. Schoelkopf}
\affiliation{Departments of Applied Physics and Physics, Yale University, New Haven, Connecticut 06511, USA}

\author{S. M. Girvin}
\affiliation{Departments of Applied Physics and Physics, Yale University, New Haven, Connecticut 06511, USA}

\author{Liang Jiang}
\affiliation{Departments of Applied Physics and Physics, Yale University, New Haven, Connecticut 06511, USA}

\begin{abstract}
Hybrid quantum systems in which acoustic resonators couple to superconducting qubits are promising quantum information platforms.
High quality factors and small mode volumes make acoustic modes ideal quantum memories, while the qubit-phonon coupling enables the initialization and manipulation of quantum states. 
We present a scheme for quantum computing with multimode quantum acoustic systems, and based on this scheme, propose a hardware-efficient implementation of a quantum random access memory (QRAM). 
Quantum information is stored in high-Q phonon modes, and couplings between modes are engineered by applying off-resonant drives to a transmon qubit.
In comparison to existing proposals that involve directly exciting the qubit, 
this scheme can offer a substantial improvement in gate fidelity for long-lived acoustic modes. 
We show how these engineered phonon-phonon couplings can be used to access data in superposition according to the state of designated address modes---implementing a QRAM on a single chip. 
\end{abstract}
\maketitle


\emph{Introduction.---}The coupling of superconducting qubits to microwave resonators, termed circuit quantum electrodynamics (cQED)~\cite{blais2007,schoelkopf2008}, constitutes one of today's most promising quantum computing architectures. 
Microwave modes provide good quantum memories~\cite{reagor2016}, while superconducting nonlinearities enable the initialization~\cite{hofheinz2008}, manipulation~\cite{krastanov2015,heeres2017}, readout~\cite{sun2014}, and protection~\cite{ofek2016,hu2019} of quantum states encoded in microwave photons. However, long microwave wavelengths could potentially limit the scalability of cQED systems. On-chip resonators face trade-offs between compactness and quality factor~\cite{geerlings2012,wenner2011}, and microwave modes with millisecond coherence have thus far only been demonstrated in large 3D cavities~\cite{reagor2016,romanenko2018}.

\begin{figure}[b]
\includegraphics[width=1\columnwidth]{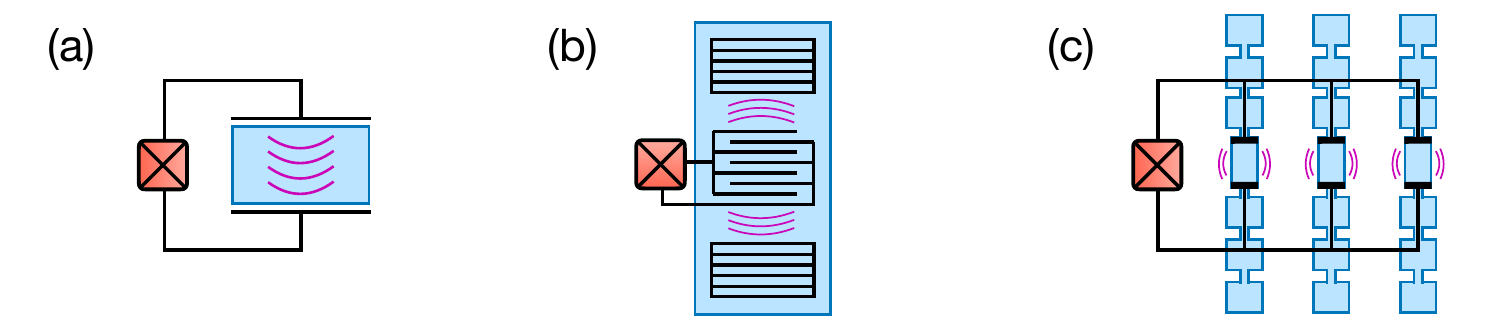}\caption{Multimode cQAD. A transmon qubit (red) is piezoelectrically coupled
to (a) a bulk acoustic wave resonator, (b) a surface acoustic wave resonator, or (c) an array of phononic crystal resonators. \label{fig:cQAD} }
\end{figure}

Recently, coherent couplings between superconducting qubits and acoustic resonators have been demonstrated in a remarkable series of  experiments~\cite{oconnell2010,pirkkalainen2013,gustafsson2014,chu2017,chu2018,kervinen2018,manenti2017,noguchi2017,satzinger2018,moores2018,bolgar2018,sletten2019a,arrangoiz-arriola2019}.
These so-called circuit quantum acoustodynamic (cQAD) systems (Fig.~\ref{fig:cQAD}) possess many of the  advantageous properties of cQED systems, e.g., superconducting qubits can be used to generate arbitrary superpositions of acoustic Fock states~\cite{chu2018,satzinger2018} and perform phonon-number resolving measurements~\cite{arrangoiz-arriola2019,sletten2019a}. Yet relative to electromagnetic modes, acoustic modes can provide dramatic benefits in terms of size and coherence times. The velocities of light and sound differ by five orders of magnitude, and short acoustic wavelengths enable the fabrication of ultra-compact phononic resonators~\cite{safavi-naeini2019}. Furthermore, acoustic modes can be exceptionally well-isolated from their environments---quality factors in excess of $10^{10}$ were recently demonstrated in GHz frequency phononic crystal resonators~\cite{maccabe2019}.
A variety of applications for such platforms have been proposed, including quantum transduction~\cite{schuetz2015}, entanglement generation~\cite{cleland2004,bienfait2019}, and quantum signal processing~\cite{guo2017,andersson2018}, but surprisingly the direct use of cQAD systems for quantum computing has received relatively little attention, with the notable exception of Ref.~\cite{pechal2018}.

In this work, we propose a hardware-efficient and scalable quantum computing architecture for multimode cQAD systems. 
Quantum information is stored in high-quality acoustic modes, 
and interactions between modes are engineered by applying off-resonant drives to an ancillary superconducting transmon qubit.
During these operations, the transmon is only virtually excited, so the effects of transmon decoherence are mitigated. This is a crucial property, since the transmon's decoherence rate can exceed that of the phonons by orders of magnitude.
In comparison to existing proposals that involve directly exciting the transmon~\cite{naik2017,pechal2018}, this virtual approach can offer substantial improvement in gate fidelity for long-lived phonons. This scheme is also directly applicable to multimode cQED~\cite{naik2017}. 

Furthermore, to demonstrate the benefits that the proposed cQAD architecture affords in hardware efficiency, we propose an implementation of a quantum random access memory (QRAM)~\cite{giovannetti2008,giovannetti2008a}. A classical RAM is a device that can query a database. Given an address $j$ as input, the RAM outputs the element $D_{j}$ stored at position $j$ in the database. 
Analogously, a QRAM is a device that, when provided with a superposition of addresses, returns a correlated superposition of data,
\begin{equation}
\sum_{j=1}^N\alpha_{j}\ket{j}_{a}\ket{0}_{b}\xrightarrow[]{\text{QRAM}}\sum_{j=1}^N\alpha_{j}\ket{j}_{a}\ket{D_{j}}_{b},
\label{eq:QRAM_def}
\end{equation}
where the subscripts $a$ and $b$ denote the address and output qubit registers, respectively. 
The ability to perform such queries efficiently, i.e.~in $\log N$ time, is a prerequisite for a variety of quantum algorithms that provide speedups over their classical counterparts~\cite{grover1997,harrow2009,biamonte2017}. 
However, building a QRAM is highly non-trivial; even a small-scale QRAM of the sort described in Ref.~\cite{giovannetti2008} has yet to be experimentally demonstrated. 
One major challenge is that, to query a database of size $N$, a QRAM requires $O(N)$ quantum resources~\cite{giovannetti2008a}. 
Hardware-efficiency is thus crucial for QRAM queries of large datasets, and in our implementation this efficiency is enabled by the on-chip integration of superconducting circuits with compact acoustic resonators.
Out proposal both reduces physical resource requirements and provides a roadmap for a near-term demonstration of QRAM.

Note that ``random access quantum memories''~\cite{mariantoni2011,jiang2019,naik2017}---memories that can take a single $\emph{classical}$ address $j$ as input and return a corresponding qubit $\ket{\psi_j}$---cannot generally perform operation (1) and should not be confused with QRAM, which can use superpositions of addresses to query multiple memory elements.


\emph{Quantum computing in cQAD.---}In multimode cQAD, a transmon qubit is piezoelectrically coupled to a collection of acoustic modes. These modes can be supported in 
bulk acoustic wave (BAW)~\cite{chu2017,chu2018,kervinen2018} or surface acoustic wave (SAW)~\cite{manenti2017,noguchi2017,satzinger2018,moores2018,bolgar2018,sletten2019a} resonators, or in an array of phononic crystal (PC) resonators~\cite{arrangoiz-arriola2019} (Fig.~\ref{fig:cQAD}). 
Quality factors of $\approx 10^{5}$, $10^{8}$, and $10^{10}$ have been measured at GHz frequencies in SAW~\cite{manenti2016,aref2016a}, BAW~\cite{renninger2018,kharel2018a}, and PC resonators~\cite{maccabe2019}, respectively, and 
the transmon can be simultaneously coupled to many high-Q modes on a single chip~\cite{chu2017}.
These systems can be described by the Hamiltonian 
\begin{equation}
\begin{split}
H &=\omega_{q}q^{\dagger}q-\frac{\alpha}{2}q^{\dagger}q^{\dagger}qq\\
&+\sum_{k}\left(\omega_{k}m_{k}^{\dagger}m_{k}+g_k q^{\dagger}m_{k}+ g_k^* q m_{k}^{\dagger} \right)+H_{d}\label{eq:cQAD_Hamiltonian}
\end{split}
\end{equation}
Here, $q$ and $m_k$ denote the annihilation operators for the transmon and phonon modes, respectively. The transmon is modeled as an anharmonic oscillator with Kerr nonlinearity $\alpha$ and is coupled to the $k^{th}$ phonon mode with strength $g_k$ (typically a few MHz~\cite{manenti2017,arrangoiz-arriola2019,han2016}). In combination with external drives on the transmon $H_d = \sum_j \Omega_j q^\dagger e^{-i\omega_j t}+ \mathrm{H.c.}$, this coupling provides the basic tool to initialize, manipulate, and measure phononic qubits~\cite{chu2018,satzinger2018}. For example, itinerant photon-encoded qubits sent to the system can be routed into a particular phonon mode via pitch-and-catch schemes~\cite{palomaki2013,pechal2014,srinivasan2014,axline2018a,kurpiers2018}.

\begin{figure}[tbp]
\centering{}\includegraphics[width=1.0\columnwidth]{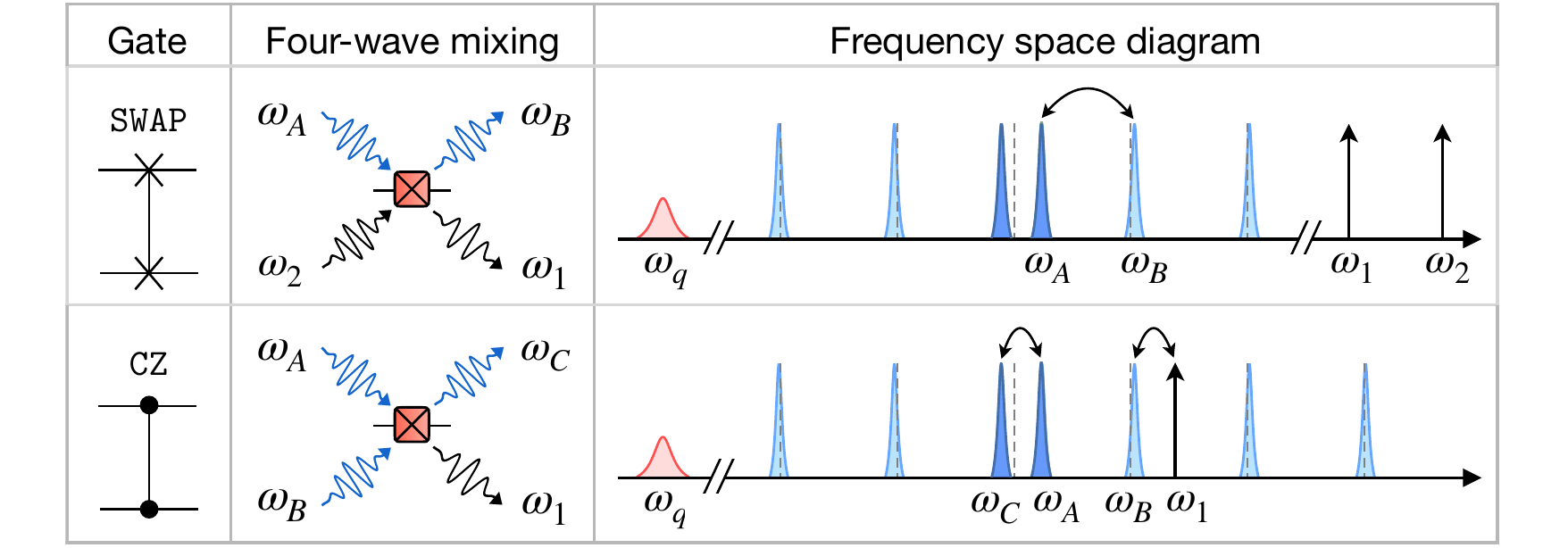}\caption{\label{fig:phonon_gates}Phonon-phonon gates. \texttt{SWAP}: Applying two drives with $\omega_2-\omega_1=\omega_B-\omega_A$ creates an effective coupling between modes $A$ and $B$.  \texttt{CZ}: Applying a single drive with $\omega_1=\omega_A+\omega_B-\omega_C$ creates an effective three-mode coupling between modes $A$, $B$, and $C$.  Frequency shifts of strongly hybridized modes (dark blue) can enable selective coupling when the modes are otherwise uniformly spaced (dashed lines denote uniform spacing)~\cite{SM}.
}
\end{figure}

Interactions between phonon modes can be engineered by applying off-resonant drives to the transmon, and we use these interactions to implement a universal gate set for phononic qubits. 
The main idea is that the transmon's Kerr nonlinearity enables it to act as a four-wave mixer~\cite{mutus2013,leghtas2015,gao2018,zhang2019}, so phonons can be converted from one frequency to another by driving the transmon.
For example, phonons can be converted from frequency $\omega_{A}$ to $\omega_{B}$ by applying two drive tones whose frequencies $\omega_{1,2}$ satisfy the resonance condition $\omega_{2}-\omega_{1}=\omega_{B}-\omega_{A}$, see Fig.~\ref{fig:phonon_gates}. This driving gives rise to an effective Hamiltonian  $H=g_{v}^{(1)}m_{A}m_{B}^{\dagger}+\mathrm{H.c.}$, where $g_{v}^{(1)}=-2\alpha\frac{g_A}{\delta_{A}}\frac{g_B^{*}}{\delta_{B}}\frac{\Omega_{1}^{*}}{\delta_{1}}\frac{\Omega_{2}}{\delta_{2}}(1-\beta^{(1)})$. 
Here, $\delta_j \equiv \omega_j - \omega_q$, and $\beta^{(1)}$ is a drive-dependent correction (see supplementary material~\cite{[{}][{See Supplementary Material.}]SM} for derivations). 
Evolution under this coupling for a time $\pi/2g_{v}^{(1)}$ implements a \texttt{SWAP} gate, which exchanges the states of modes $m_{A}$ and $m_{B}$, while evolution for a time $\pi/4g_{v}^{(1)}$ implements a 50:50 beamsplitter operation~\cite{gao2018}.

Three-mode interactions can be similarly engineered (Fig.~\ref{fig:phonon_gates}). 
Applying a single drive tone with frequency $\omega_{1}=\omega_{A}+\omega_{B}-\omega_{C}$ gives rise to the effective Hamiltonian $H=g_{v}^{(2)}m_{A}m_{B}m_{C}^{\dagger}+\mathrm{H.c.}$, where $g_{v}^{(2)}=-2\alpha\frac{g_A}{\delta_{A}}\frac{g_B}{\delta_{B}}\frac{g_C^{*}}{\delta_{C}}\frac{\Omega_{1}^{*}}{\delta_{1}}(1-\beta^{(2)})$~\cite{SM}. 
This three-mode interaction can be used to implement a controlled phase (\texttt{CZ}) gate for qubits encoded in the $\ket{0,1}$ phonon Fock states~\cite{langford2011}. 
To perform a \texttt{CZ} gate between qubits in modes $A$ and $B$, mode $C$ is used as an ancilla and initialized in $\ket{0}$. 
Evolving for a time $\pi/g_{v}^{(2)}$ then enacts the mapping $\ket{110}_{ABC}\rightarrow\ket{001}\rightarrow -\ket{110}$, while leaving all other initial states unaffected. 
The state $\ket{11}_{AB}$ acquires a relative geometric phase, thereby implementing the \texttt{CZ} gate. 

A variety of other operations can be similarly implemented. Single- and two-mode squeezing can be implemented by driving the transmon at appropriate frequencies, and phase shifts can be imparted by tuning the drive phases during \texttt{SWAP} operations. Together, these two- and three-mode interactions are universal~\cite{niu2018,SM}. In the remainder of this work, however, we focus on the beamsplitter, \texttt{SWAP}, and \texttt{CZ} operations, as these are the only operations we require to implement a QRAM.

Note that in BAW and SAW resonators, phonon mode frequencies are approximately uniformly spaced, i.e.~$\omega_{j+1}-\omega_j=\nu$, where $\nu$ is the free spectral range. 
This uniform spacing can lead to problematic degeneracies in the resonance conditions above. Nonuniform mode spacing is thus necessary to enable selective coupling, and in~\cite{SM} we describe several ways to engineer nonuniformity in BAW and SAW systems.
As shown in Fig.~\ref{fig:phonon_gates}, one approach is to couple the phonons to an external mode, such as a microwave resonator, so that the resulting hybridization shifts mode frequencies~\cite{han2016}. 
In~\cite{SM}, we also introduce a metric, $\Delta\nu$, to quantify the nonuniformity. Roughly speaking, $\Delta\nu$ is the scale at which the mode spacing varies.


\emph{Gate fidelities.---}During the gates described above, the transmon is never directly excited; instead, it is only \emph{virtually} excited, so infidelity attributable to transmon decoherence is suppressed.
These virtual gates can thus provide great advantage in cQAD systems, where transmon decoherence is likely to be the limiting factor. 
This is in contrast to existing proposals~\cite{naik2017,pechal2018}, in which gates between resonator mode qubits are implemented by swapping information \emph{directly} into the transmon using resonant interactions of the form $g_{d}(q^{\dagger}m+qm^{\dagger})$, which can be engineered, e.g., by modulating the transmon's frequency. 
In the following, we compare the predicted fidelities of the \emph{virtual} gates proposed here and the \emph{direct} gates considered in Refs.~\cite{naik2017,pechal2018}.

In a multimode architecture, there exists a fundamental tradeoff between decoherence and spectral crowding. Slower gates are more prone to decoherence, while faster gates have reduced frequency resolution and can disrupt other modes. 
The infidelities of the direct and virtual gates, respectively $1-\mathcal{F}_{d}$ and $1-\mathcal{F}_{v}$, can be approximated as a sum of contributions from these two effects~\cite{pechal2018,SM},
\begin{align}
1-\mathcal{{F}}_{d}&\approx (\kappa+\gamma)\frac{c_{d}\pi}{2g_{d}}+\left(\frac{g_{d}}{\nu}\right)^{2},\label{eq:F_d} \\
1-\mathcal{{F}}_{v}&\approx \bar{\kappa}_{\gamma}\frac{c_{v}\pi}{2g_{v}}+\left(\frac{g_{v}}{\Delta\nu}\right)^{2},\label{eq:F_v}
\end{align}
where $\kappa$ and $\gamma$ are the bare phonon and transmon decoherence rates, and $c_{d,v}$ are constants accounting for the durations of each gate ($c_{v}=1$ for \texttt{SWAP}, and $c_{v}=2$ for \texttt{CZ}, as these gates have durations $\pi/2g_v$ and $\pi/g_v$ respectively. As discussed in Ref.~\cite{naik2017}, $c_{d}=5$ for \texttt{SWAP} and $c_{d}=4$ for \texttt{CZ}.) 

The first terms in Eq.~(\ref{eq:F_d}) and Eq.~(\ref{eq:F_v}) account for decoherence. During direct gates, information spends roughly equal time in the phonon and transmon modes, so the total decoherence rate is $\kappa+\gamma$. 
During virtual gates, the total decoherence rate, $\bar{\kappa}_{\gamma}$, is $\bar{\kappa}_{\gamma}=(\kappa_{\gamma}^{A}+\kappa_{\gamma}^{B})$ for \texttt{SWAP}, and $\bar{\kappa}_{\gamma}=(\kappa_{\gamma}^{A}+\kappa_{\gamma}^{B}+\kappa_{\gamma}^{C})/2$ for \texttt{CZ}.
Here, $\kappa_{\gamma}^{j}=\kappa+\gamma(g_j/\delta_{j})^{2}(1+\beta^{(\gamma)})$ denotes the dressed decay rate of mode $j$, which includes a contribution from the inverse Purcell effect~\cite{reagor2016,zhang2019} and a drive-dependent correction $\beta^{(\gamma)}$~\cite{SM}. The second term in each expression accounts for spectral crowding. 
The probability of accidentally exciting another mode scales as $(g_{d}/\nu)^{2}$ in the direct case, and as $(g_{v}/\Delta\nu)^{2}$ in the virtual case. We note that the infidelities~(\ref{eq:F_d},\ref{eq:F_v}) are defined with respect to the ideal \emph{two-qubit} unitaries; see~\cite{SM} for derivations. 

The competition between decoherence and spectral crowding results in an optimal coupling rate~\cite{pechal2018}. 
By adjusting the drive strengths, $g_{d,v}$ can be tuned to their respective optima. The optimal infidelities are 
\begin{align}
1-\mathcal{F}_{d}  &\approx\frac{3}{2} \left[\frac{ c_d \pi(\kappa+\gamma)}{\sqrt{2}\nu}\right]^{2/3},\\
1-\mathcal{{F}}_{v}  &\approx \frac{3}{2} \left[\frac{c_v \pi \bar{\kappa}_{\gamma}}{\sqrt{2}\Delta\nu}\right]^{2/3}.
\end{align}
While transmon and phonon decoherence contribute equally to $1-\mathcal{F}_d$, transmon decoherence only makes a small contribution to $1-\mathcal{F}_v$ via the inverse Purcell effect, wherein $\gamma$ is suppressed by a factor of $(g/\delta)^2\ll1$. The virtual gates can thus be expected to attain higher fidelities when there is a large disparity between $\gamma$ and $\kappa$, i.e.~for sufficiently long-lived phonon modes. Indeed, $\mathcal{F}_{v}>\mathcal{F}_{d}$ whenever $\kappa_{\gamma}\lesssim(\kappa+\gamma)\Delta\nu/\nu$, provided the optimal coupling rates can be reached.

\begin{figure}[tbp]
\centering{}\includegraphics[width=1.0\columnwidth]{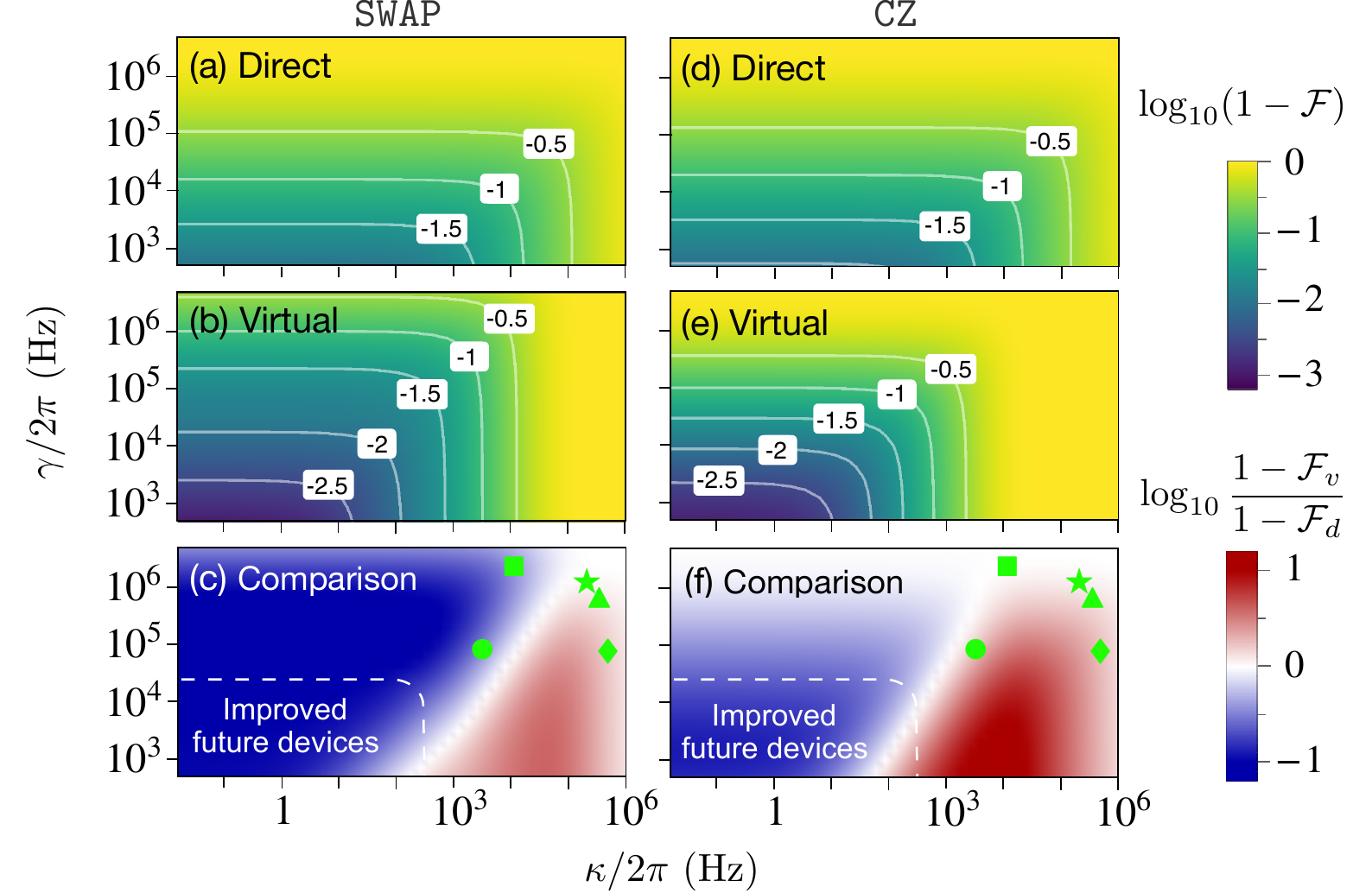}
\caption{Comparison of direct and virtual operations.
(a,b) $\log_{10}(1-\mathcal{F})$ for the direct and virtual \texttt{SWAP}
operations, respectively. The couplings are optimized subject to constraints ($g_d \in [0,g]$, constraints on $g_v$ are discussed in~\cite{SM}). (c) Comparison of direct and virtual \texttt{SWAP}
operations. The log ratio of the infidelities is plotted, with the
virtual operations attaining higher fidelities in the blue region. (d,e) Log$_{10}$
infidelity for the direct and virtual \texttt{CZ} operations.
(f) Comparison of \texttt{CZ} operations. 
For reference, the symbols \{\tikzcircle{2.5pt},$\blacksquare$,$\blacktriangle$,$\blacklozenge$,$\bigstar$\} respectively denote the decoherence rates $\kappa$ (phonon) and $\gamma$ (transmon) measured in Refs.~\cite{chu2018},~\cite{manenti2017},~\cite{arrangoiz-arriola2019},~\cite{satzinger2018}, and~\cite{moores2018}.
Note, however, that
the plots are generated using typical parameter values, not specific values from any one experiment.
Parameters: $g/2\pi=10\si{\mega\hertz},$ $\delta/2\pi=100\si{\mega\hertz}$,
$\nu/2\pi=10\si{\mega\hertz}$, and $\Delta\nu/2\pi=1\si{\mega\hertz}$.  \label{fig:direct_virtual_comparison}
}
\end{figure}

In Fig.~\ref{fig:direct_virtual_comparison}, we plot the optimal infidelities of direct and virtual gates as a function of $\kappa$ and $\gamma$ for currently feasible experimental parameters. 
The comparison reveals that virtual gates can be performed with high fidelity ($>$99\%) given long-lived phonons, and that virtual gates attain higher fidelities than direct gates in the same regime. 
Indeed, realistic improvements in phonon coherence are likely to bring near-term devices into this $\mathcal{F}_v \gg \mathcal{F}_d$ regime (Fig.~\ref{fig:direct_virtual_comparison}c,f). 

We briefly note other factors relevant to the comparison of direct and virtual gates.
\emph{Multi-phonon encodings.} Direct gates require that qubits be encoded in the $\ket{0,1}$ phonon Fock states, while virtual operations are compatible with multi-phonon encodings, including some bosonic quantum error-correcting codes~\cite{niu2018, niu2018a}. 
\emph{Parallelism.} Direct gates must be executed serially, while virtual gates can be executed in parallel by simultaneously applying the requisite drives (though care should be taken to ensure that the additional drives do not bring spurious couplings on resonance).
\emph{Speed.} Virtual gates are inherently slower than direct gates, with realistically attainable coupling rates $g_v/2\pi \sim 10-\SI{100}{\kilo\hertz}$~\cite{SM}. 


\emph{QRAM Implementation.---}To illustrate the advantages of cQAD systems, we propose an implementation of a QRAM~\cite{giovannetti2008,giovannetti2008a}. As defined by~Eq.(\ref{eq:QRAM_def}), a QRAM is a device which can query a database with an address in superposition. 
The ability to perform such queries efficiently is a prerequisite for a variety of quantum algorithms, including Grover's search~\cite{grover1997}, matrix inversion~\cite{harrow2009}, and various proposals in the field of quantum machine learning~\cite{biamonte2017}. 
While demanding hardware and connectivity requirements have thus far precluded an experimental demonstration of a QRAM, our proposed cQAD implementation is naturally hardware-efficient. Indeed, a small-scale cQAD QRAM can be implemented with a single multimode resonator.

\begin{figure}[tbp]
\includegraphics[width=1.0\columnwidth]{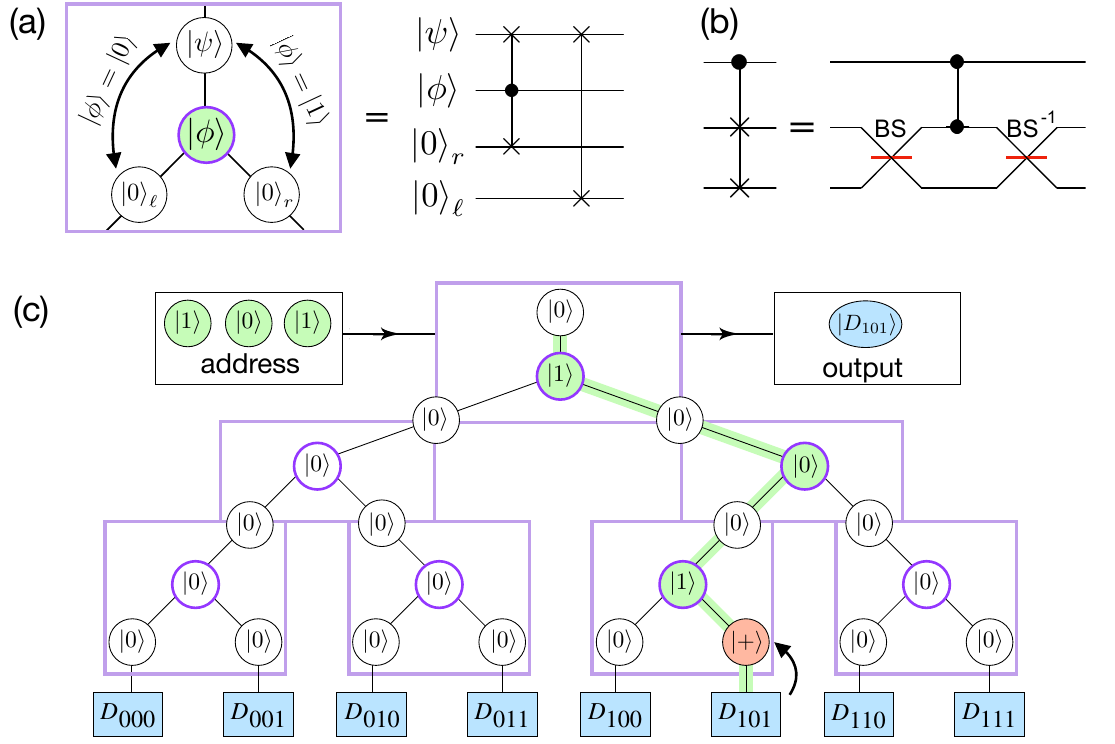}
\caption{cQAD implementation of QRAM. 
(a) Quantum router. 
Each circle represents a phonon mode. The router directs the qubit $\ket{\psi}$ in the incoming mode (top) to either the right or left mode conditioned on the state of the routing qubit $\ket{\phi}$. 
(b) Controlled \texttt{SWAP}. Note that this circuit implements the gate only within the relevant subspace of $<2$ total phonons in the modes to be swapped.
(c) QRAM implementation. 
Address qubits (green) are routed into position one-by-one, carving out a path to the database. 
The bus qubit (red) follows this path to retrieve the data $D_j$. 
The bus and address qubits are then routed back out of the tree to complete the query. 
The database (blue squares) can be either classical or quantum. 
In the former case, the bus is initially prepared in $\ket{+}$, and classical bits are copied to the bus by applying phase shifts to each mode at the bottom of the tree. 
In the latter case, the data qubit is extracted through a sequence of controlled \texttt{SWAP} operations.
See~\cite{SM} for details. \label{fig:phonon_QRAM}}
\end{figure}

The elementary building block of our QRAM implementation is a \emph{quantum router}, shown in Fig.~\ref{fig:phonon_QRAM}(a). 
The router directs an incoming qubit into different output modes conditioned on the state of a routing qubit. When the routing qubit is in state $\ket{0}(\ket{1})$, an incoming qubit $\ket{\psi}$ in the top mode is swapped to the left(right) mode. The routers are implemented using the operations described above: the routing circuit contains a \texttt{SWAP} and a controlled-\texttt{SWAP} gate, the latter of which is implemented using \texttt{CZ} and beamsplitter operations (Fig.~\ref{fig:phonon_QRAM}b).  

To implement a QRAM, a collection of routers is arranged in a binary tree, with the outputs of routers at one level acting as inputs to routers at the next (Fig.~\ref{fig:phonon_QRAM}c). To query the database at the bottom of the tree, qubits from the address register are routed sequentially into the tree, with earlier address qubits controlling the routing of later ones in a ``bucket-brigade'' scheme~\cite{giovannetti2008}. 
A ``bus'' qubit then follows the path paved by the address qubits and extracts the data, after which it is routed back out of the tree and into the output register.
Finally, to disentangle the address and routers, the address qubits are routed back out of the tree. 
Since all routing operations are quantum-controlled, preparing the address register in superposition allows access to the data in superposition, thereby implementing operation~(\ref{eq:QRAM_def}). Further details are provided in~\cite{SM}.

We highlight three appealing properties of this cQAD\nobreakdash-based QRAM.
\begin{enumerate}[wide, labelwidth=!, labelindent=0pt,topsep=0pt,itemsep=-1ex]
    \item \emph{Hardware-efficiency.} Hundreds of phonon modes can simultaneously couple to a transmon on a single chip~\cite{chu2017}. Thus, the hardware and fabrication cost of a cQAD-based QRAM can be drastically reduced in comparison to cavity-~\cite{giovannetti2008a} or circuit-QED~\cite{kyaw2015} implementations. 
    \item \emph{Scalability.}  
    It is not necessary to control all routing through a single transmon; since only adjacent routers are coupled, different regions of the tree can be controlled and implemented independently. 
    For example, the QRAM can be built out of several modules, where each module comprises a group of routers controlled by one transmon. 
    The phononic modes in each module could be supported in physically separate resonators, or multiple transmons could be simultaneously coupled to the same multimode resonator to give access to a large bandwidth of modes, potentially spanning several GHz~\cite{chu2017}. 
    \item \emph{Error resilience.} Because our implementation follows the bucket-brigade model, it inherits a favorable $\log N$ error scaling~\cite{giovannetti2008,giovannetti2008a,arunachalam2015}. In particular, the scaling argument of Ref.~\cite{giovannetti2008} directly applies to phonon-loss errors: the query infidelity scales as $1-\mathcal F \sim \varepsilon \log N$, where $\varepsilon$ is the phonon loss probability. Remarkably, one can show that the infidelity scales logarithmically for arbitrary independent, incoherent errors, such as phonon loss, dephasing, and heating~\cite{[{}][{C.~T.~Hann and L.~Jiang, (in preparation).}]hanninpreparation}. 
\end{enumerate}

\emph{Discussion.---}We have proposed a quantum computing architecture for multimode cQAD and an implementation of a QRAM based on it. 
The implementation is hardware-efficient, owing to the compactness of multimode cQAD systems that is enabled by small acoustic wavelengths. We emphasize that hardware efficiency is not only crucial for scaling to large system sizes, but that it is also particularly advantageous for near-term experiments. Indeed, a small-scale QRAM can be implemented even with just a single multimode resonator. In the long term, the use of bosonic quantum error correcting codes~\cite{albert2018,niu2018a} and compatible logical gates~\cite{lau2016,niu2018,gao2019} to implement a QRAM that is both fault-tolerant and hardware-efficient is an intriguing direction for future research. These ideas can also be directly applied to multimode cQED.

To be viable, our scheme requires long phonon coherence times ($1/\kappa \gg 1/g_v$). 
Though loss due to intrinsic material processes like phonon-phonon scattering or two level systems limit phonon coherence, such mechanisms should not prevent access to this regime. 
Indeed, both BAW and PC quality factors can approach $10^{10}$ before encountering such limits~\cite{goryachev2012,galliou2013,maccabe2019}, corresponding to $\kappa/2\pi\sim 1$Hz at GHz frequencies.
Additionally, intrinsic sources of phonon dephasing are not expected~\cite{faust2013}. 
Phonon decoherence in current cQAD experiments is thus likely dominated by extrinsic mechanisms that can be mitigated with improved fabrication techniques, though the extent to which the coupling to superconducting circuits may limit phonon coherence in cQAD is an important open question. 

\emph{Acknowledgements.---}We thank Vijay Jain, Prashanta Kharel, Patricio Arrangoiz-Arriola, Alex Wollack, Hong Tang, Peter Rakich, and Oskar Painter for helpful discussions. C.T.H. acknowledges support from the NSF Graduate Research Fellowship Program (DGE1752134). We acknowledge support from the ARL-CDQI (W911NF-15-2-0067, W911NF-18-2-0237), ARO (W911NF-18-1-0020, W911NF-18-1-0212), ARO MURI (W911NF-16-1-0349), AFOSR MURI (FA9550-15-1-0015), DOE (DE-SC0019406), NSF (EFMA-1640959, DMR-1609326), and the Packard Foundation (2013-39273).

\normalbaselines
\bibliographystyle{apsrev4-1}
\let\oldaddcontentsline\addcontentsline
\renewcommand{\addcontentsline}[3]{}
\bibliography{phonon_references}
\let\addcontentsline\oldaddcontentsline


\widetext
\setcounter{equation}{0}
\setcounter{figure}{0}
\setcounter{table}{0}
\makeatletter
\renewcommand{\theequation}{S\arabic{equation}}
\renewcommand{\thefigure}{S\arabic{figure}}

\newpage
\begin{center}
\vspace{2em}
{\large\bf Supplemental Material: Hardware-efficient quantum random access memory with hybrid quantum acoustic systems}
\end{center}
\vspace{-2em}
\tableofcontents

\section{Virtual coupling rates}
In this section, we study the virtual coupling rates
\begin{align}
g_{v}^{(1)} & =-2\alpha\xi_{1}^{*}\xi_{2}\lambda_{A}\lambda_{B}^{*}(1-\beta^{(1)})\label{eq:gv1_expression},\\
g_{v}^{(2)} & =-2\alpha\xi_{1}^{*}\lambda_{A}\lambda_{B}^{*}\lambda_{C}(1-\beta^{(2)}).\label{eq:gv2_expression}
\end{align} 
Below, we define the notation, derive these expressions, and discuss the importance of the corrections $\beta^{(1,2)}$ for cQAD systems. Then, in order to verify the accuracy of these expressions, we compare them to numerical results obtained using the Floquet theory methods of Ref.~\cite{zhang2019}.  

\subsection{Derivation of the virtual coupling rates}
To derive the expressions~(\ref{eq:gv1_expression}) and (\ref{eq:gv2_expression}), we begin with the multimode cQAD Hamiltonian (Eq.~(2) of the main text) and perform a unitary transformation defined by $U_1 = \exp i H_0 t$, where $H_0 = \omega_q q^\dagger q+\sum_k \omega_k m_k^\dagger m_k$. Thus,
\begin{equation}
H=\sum_{j}\left(\Omega_{j}q^{\dagger}e^{-i\delta_{j}t}+\mathrm{H.c.}\right)+\sum_{k}\left(g_km_{k}q^{\dagger}e^{-i\delta_{k}t}+\mathrm{H.c.}\right)-\frac{\alpha}{2}q^{\dagger}q^{\dagger}qq,
\end{equation}
where $\delta_k = \omega_k-\omega_q$ is the detuning of the $k^{th}$ phonon mode, while $\delta_j = \omega_j-\omega_q$ and $\Omega_j$ are the detuning and the strength of the $j^{th}$ drive tone, respectively.
In the spirit of Ref.~\cite{nigg2012},
we first perform unitary transformations to eliminate the qubit-phonon
couplings and drive terms then consider the effects of the anharmonicity. For notational convenience, we introduce the dimensionless
parameters $\lambda_{k}\equiv g_k/\delta_{k}$ and $\xi_{j}\equiv\Omega_{j}/\delta_{j}$.
To leading order in $\lambda_{k}\ll1$, the unitary that eliminates
the couplings is $U_{2}=\exp\sum_{k}(\lambda_{k}^{*}m_{k}^{\dagger}qe^{i\delta_{k}t}-H.c)$,
and that which eliminates the drives is $U_{3}=\exp\sum_{j}(\xi_{j}^{*}qe^{i\delta_{j}t}-H.c)$.
The combined effect of these two transformations is to enact the mapping
$q\rightarrow q+\sum_{j}\xi_{j}e^{-i\delta_{j}t}+\sum_{k}\lambda_{k}m_{k}e^{-i\delta_{k}t}\equiv Q,$
so that the Hamiltonian becomes
\begin{equation}
H=-\frac{\alpha}{2}Q^{\dagger}Q^{\dagger}QQ.\label{eq:Q_Hamiltonian}
\end{equation}
Note that we have neglected linear terms of the form
$(\Omega_{j}^{*}\lambda_{k}m_{k}e^{i(\delta_{j}-\delta_{k})t}+\mathrm{H.c.})$. This omission is justified in the RWA provided that $|\delta_{j}-\delta_{k}|\gg\lambda_{k}\Omega_{j}$,
i.e.~that the drives are sufficiently far detuned from any modes in
which we are interested.
For the moment, we also neglect frequency (Stark) shifts of the phononic eigenmodes---we consider their effects in Sec.~(\ref{frequency_shifts}). 

When two drive tones are applied whose frequencies satisfy the resonance condition $\omega_{2}-\omega_{1}=\omega_{B}-\omega_{A}$,
the Hamiltonian~(\ref{eq:Q_Hamiltonian}) contains a resonant beamsplitter-type coupling, $g_{v}^{(1)}m_{A}m_{B}^{\dagger}+\mathrm{H.c.},$
where
\begin{equation}
g_{v}^{(1)}=-2\alpha\xi_{1}^{*}\xi_{2}\lambda_{A}\lambda_{B}^{*}.\label{eq:BB_1}
\end{equation}
Similarly, when a single drive tone is applied with frequency\footnote{In this section we consider the case $\omega_A < \omega_B < \omega_1 < \omega_C$, which nicely highlights the similarities between $g_v^{(1)}$ and $g_v^{(2)}$. The derivations proceed analogously for other cases, such as the case of $ \omega_C<\omega_A < \omega_B < \omega_1$ shown in Fig.~2 of the main text.} $\omega_{1}=\omega_{A}+\omega_{C}-\omega_{B}$,
the Hamiltonian contains a resonant three-mode coupling
$g_{v}^{(2)}m_{A}m_{B}^{\dagger}m_{C}+\mathrm{H.c.},$
where
\begin{equation}
g_{v}^{(2)}=-2\alpha\xi_{1}^{*}\lambda_{A}\lambda_{B}^{*}\lambda_{C}.\label{eq:BB_2}
\end{equation}


\subsection{Corrections to the virtual coupling rates}

The Hamiltonian (\ref{eq:Q_Hamiltonian}) contains many terms beyond
just the resonant terms discussed above (see Table~\ref{tab:ham_terms}). 
Most of these terms are rapidly-rotating and can be neglected in the RWA assuming dispersive coupling
($\lambda\ll1$) and weak drives $(\xi\ll1)$.
However, other terms can produce corrections $\beta^{(1,2)}$
to the coupling rates.
In this section, we first calculate these corrections to leading order
in $\lambda$ and $\xi$. Then, we derive nonperturbative contributions
associated with the AC Stark shift. 

\def\arraystretch{1.4}
\begin{table}[tbph]
\caption{Catalog of terms in the Hamiltonian (\ref{eq:Q_Hamiltonian}). Summations run over all drives and all modes, including the transmon mode $q$, for which $\lambda_{q}=1$ and $\delta_{q}=0$. \label{tab:ham_terms}
}
\centering{}
\begin{tabular}{|c|c|}
\hline 
Term & Description  \tabularnewline
\hline 
$\frac{\alpha}{2}\sum_{i,j,k,l}\xi_{i}^{*}\xi_{j}^{*}\xi_{k}\lambda_{l}m_{l}e^{i(\delta_{i}+\delta_{j}-\delta_{k}-\delta_{l})t}+\mathrm{H.c.}$ 
&  Drive
\tabularnewline
\hline 
$\frac{\alpha}{2}\sum_{i,j,k,l}\xi_{i}^{*}\xi_{j}\lambda_{k}^{*}\lambda_{l}m_{k}^{\dagger}m_{l}e^{i(\delta_{i}-\delta_{j}+\delta_{k}-\delta_{l})t}+\mathrm{H.c.}$ 
& Beamsplitter 
\tabularnewline
\hline 
$\frac{\alpha}{2}\sum_{i,j,k,l}\xi_{i}^{*}\xi_{j}^{*}\lambda_{k}\lambda_{l}m_{k}m_{l}e^{i(\delta_{i}+\delta_{j}-\delta_{k}-\delta_{l})t}+\mathrm{H.c.}$ 
& Two-mode squeezing
\tabularnewline
\hline 
$\frac{\alpha}{2}\sum_{i,j,k,l}\xi_{i}^{*}\lambda_{j}^{*}\lambda_{k}\lambda_{l}m_{j}^{\dagger}m_{k}m_{l}e^{i(\delta_{i}+\delta_{j}-\delta_{k}-\delta_{l})t}+\mathrm{H.c.}$ 
&$\chi^{(2)}$ nonlinearity
\tabularnewline
\hline 
$\frac{\alpha}{2}\sum_{i,j,k,l}\lambda_{i}^{*}\lambda_{j}^{*}\lambda_{k}\lambda_{l}m_{i}^{\dagger}m_{j}^{\dagger}m_{k}m_{l}e^{i(\delta_{i}+\delta_{j}-\delta_{k}-\delta_{l})t}+\mathrm{H.c.}$ 
&$\chi^{(3)}$ nonlinearity
\tabularnewline
\hline 
\end{tabular} 
\end{table}

The leading order contribution to $\beta^{(1,2)}$ is zeroth order
in both $\lambda$ and $\xi$. The only terms in the Hamiltonian (\ref{eq:Q_Hamiltonian})
which contribute to $\beta^{(1)}$ and $\beta^{(2)}$ at this order
are, respectively, 
\begin{equation}
\left[-\alpha(
q^{\dagger2}\xi_{2}\lambda_{A}m_{A}
+q^{\dagger2}\xi_{1}\lambda_{B}m_{B})
e^{-i(\delta_{B}+\delta_{1})t}+\mathrm{H.c.}\right]
\text{ and }
\left[-\alpha(
q^{\dagger2}\lambda_{A}m_{A}\lambda_{C}m_{C}
+q^{\dagger2}\xi_{1}\lambda_{B}m_{B})e^{-i(\delta_{B }+\delta_{1})t}+\mathrm{H.c.}\right]\label{eq:label-1}
\end{equation}
The corrections from these terms can be calculated via standard perturbation
theory, 
\[
\beta{}^{(1,2)}=\frac{\alpha}{\delta_{B}+\delta_{1}+\alpha}.
\]
For the \texttt{SWAP} operation, where the drives are far-detuned, this correction
is typically negligible. However, for the \texttt{CZ} operation, this correction
can significantly reduce the coupling rate since $\delta_{1},\delta_{B}$
can be comparable to $\alpha$. We note that the expression for $\beta^{(1)}$
matches the leading order expression derived in Ref.~\cite{zhang2019}. 

Contributions to $\beta^{(1,2)}$ at higher orders in $\lambda$ can
be neglected since we have assumed the dispersive regime, $\lambda\ll1$.
Contributions at higher orders in $\xi$ can be systematically calculated
with perturbation theory in principle, but such calculations quickly
become tedious. Here, we employ an alternative approach. We consider
the AC Stark shift type terms, $-2\alpha\sum_{j}|\xi_{j}|^{2}Q^{\dagger}Q$,
and compute their contributions to $\beta^{(1,2)}$ nonperturbatively
by working in a rotating frame. 

Let $S$ denote the qubit's AC Stark shift. In the frame where the
qubit mode rotates at its Stark-shifted frequency, $\tilde{\omega_{q}}=\omega_{q}+S$,
the system Hamiltonian is
\begin{equation}
H=-Sq^{\dagger}q+\sum_{j}\left[\Omega_{j}q^\dagger e^{-i\tilde{\delta}_{j}t}+\mathrm{H.c.}\right]+\sum_{k}\left[g_k m_{k}q^{\dagger}e^{-i\tilde{\delta}_{k}t}+\mathrm{H.c.}\right]-\frac{\alpha}{2}q^{\dagger}q^{\dagger}qq
\end{equation}
where $\tilde{\delta}=\omega-\tilde{\omega_{q}}$. Performing unitary
transformations analogous to those above eliminates the coupling and
drive terms so that $H=-S\tilde{Q}^{\dagger}\tilde{Q}-\frac{\alpha}{2}\tilde{Q}^{\dagger}\tilde{Q}^{\dagger}\tilde{Q}\tilde{Q},$
where $\tilde{Q}=q+\sum_{j}\tilde{\xi}_{j}e^{-i\tilde{\delta_{j}}t}+\sum_{k}\tilde{\lambda}_{k}m_{k}e^{-i\tilde{\delta}_{k}t}$.
Here, $\tilde{\xi}_{j}=\Omega_{j}/\tilde{\delta}_{j}$ and $\tilde{\lambda}_{k}=g_k/\tilde{\delta}_{k}$.
The Stark shift terms can then be cancelled by setting\footnote{
This equation determines $S$ implicitly; to leading order in the
drives, $S=-2\alpha\sum_{j}|\Omega_{j}|^{2}/\delta_{j}^{2}$. However,
the Hamiltonian (\ref{eq:Q_Hamiltonian}) contains the terms $(\alpha\xi_{1,2}q^{\dagger2}qe^{-i\delta_{1,2}t}+\mathrm{H.c.})$,
which also contribute to $S$ at this order. Employing perturbation
theory, one finds $S=-2\alpha\sum_{j}|\Omega_{j}|^{2}/\delta_{j}(\delta_{j}+\alpha)$,
which matches the leading order calculation in Ref.~\cite{zhang2019}. This latter expression is used in the numerics throughout this work.} $S=-2\alpha\sum_{j}|\tilde{\xi}_{j}|^{2}$. 
In the frame where the Stark shift terms are eliminated, one finds
modified expressions for the corrections,
\begin{align}
\beta^{(1)} & =1-\frac{\delta_{1}\delta_{2}\delta_{A}\delta_{B}}{\tilde{\delta}_{1}\tilde{\delta}_{2}\tilde{\delta}_{A}\tilde{\delta}_{B}}\frac{\tilde{\delta}_{B}+\tilde{\delta}_{1}}{\tilde{\delta}_{B}+\tilde{\delta}_{1}+\alpha}\\
\beta^{(2)} & =1-\frac{\delta_{1}\delta_{A}\delta_{B}\delta_{C}}{\tilde{\delta}_{1}\tilde{\delta}_{A}\tilde{\delta}_{B}\tilde{\delta}_{C}}\frac{\tilde{\delta}_{B}+\tilde{\delta}_{1}}{\tilde{\delta}_{B}+\tilde{\delta}_{1}+\alpha}.
\end{align}
Hence, the coupling rates are

\begin{align}
g_{v}^{(1)} & =-2\alpha\tilde{\xi}_{1}^{*}\tilde{\xi}_{2}\tilde{\lambda}_{A}\tilde{\lambda}_{B}^{*}\frac{\tilde{\delta_{B}}+\tilde{\delta_{1}}}{\tilde{\delta_{B}}+\tilde{\delta_{1}}+\alpha}\label{eq:gv1_tilde}\\
g_{v}^{(2)} & =-2\alpha\tilde{\xi}_{1}^{*}\tilde{\lambda}_{A}\tilde{\lambda}_{B}^{*}\tilde{\lambda_{C}}\frac{\tilde{\delta_{B}}+\tilde{\delta_{1}}}{\tilde{\delta_{B}}+\tilde{\delta_{1}}+\alpha}.\label{eq:gv2_tilde}
\end{align}
These expressions have the same form as above, but with the replacements
$\delta\rightarrow\tilde{\delta}$, i.e.~detunings are now defined
relative to the qubit's Stark-shifted frequency. It follows that there also exists a Stark shift correction $\beta^{(\gamma)}$ to the inverse-Purcell enhancement 
\begin{equation}
\kappa_{\gamma}
= \kappa +\gamma (g/\delta)^2(1+\beta^{(\gamma)}),\label{eq:invPurcell_tilde}
\end{equation}
where $\beta^{(\gamma)}=(\delta/\tilde\delta)^2-1$, i.e.~$\kappa_\gamma=\kappa+\gamma(g/\tilde{\delta})^{2}$.
 These corrections are important whenever the drives are strong enough that the qubit's Stark shift becomes comparable to the drive
or mode detunings. Expressions (\ref{eq:gv1_tilde}), (\ref{eq:gv2_tilde}),
and (\ref{eq:invPurcell_tilde}) are used to produce the plots in
this work.


\subsection{Comparison with numerical Floquet calculation}

To assess the accuracy of the expressions (\ref{eq:gv1_tilde}) and
(\ref{eq:gv2_tilde}), we compare with numerical calculations of the
coupling rates using the methods developed in Ref.~\cite{zhang2019}. First, we briefly summarize the main results
of that work. The authors consider the process of engineering a bilinear
interaction between two microwave cavity modes that are mutually coupled
to a transmon qubit. Treating the couplings as a perturbation, they
calculate the linear response of the driven transmon. This perturbative
treatment is justified in the dispersive regime. They show that $g_{v}^{(1)}$
can be calculated in terms of a susceptibility matrix $\chi^{(1)}(\omega_{A},\omega_{B};\omega_{1},\omega_{2})$,
that describes the response of the driven transmon at frequency $\omega_{A}$
to a weak probe field at $\omega_{B}$, when subject to drives at
$\omega_{1}$ and $\omega_{2}$. The susceptibility can then be computed
numerically to all orders in the drive amplitudes using Floquet theory.
The authors find good quantitative agreement between their theoretical
predictions and experimental results, even for strong drives ($\xi>1$).

This approach can be directly applied to calculate $g_{v}^{(1)}$.
To calculate $g_{v}^{(2)}$, we analogously define a higher-order
susceptibility matrix $\chi^{(2)}(\omega_{A},\omega_{B},\omega_{C};\omega_{1})$
that captures the response of the transmon at frequency $\omega_{A}$
to weak probes at $\omega_{B}$ and $\omega_{C}$, when subject to
a drive at $\omega_{1}$. Rather than computing $\chi^{(2)}$ directly,
which can be numerically tedious, we note that $\chi^{(2)}$ can be
computed in terms of $\chi^{(1)}.$ In the calculation of $\chi^{(1,2)}$,
the drives and probes are treated identically at the Hamiltonian level;
both the drive and probe terms are of the form $H=f_{j}\,q^\dagger e^{-i\omega_{j}t}+\mathrm{H.c.}$.
For the drives, $f_{j}=\Omega_{j}$, while for the probes, $f_{j}=gm_{j}$.
Since the susceptibility is calculated to all orders in the drive fields
but only to leading order in the weak probe fields, going beyond
leading order does not change the result in the limit where the field
$f_{j}$ is weak. Weak probes and weak drives are thus interchangable,
the only difference being a matter of interpretation. It follows that
\begin{equation}
\chi^{(2)}(\omega_{A},\omega_{B},\omega_{C};\omega_{1})=\chi^{(1)}(\omega_{A},\omega_{B};\omega_{1},\omega_{C}).
\label{eq:chi2}
\end{equation}
This equivalence holds for $g_C\ll\delta_{C}$, which is the same limit
that was already assumed to justify the perturbative treatment. Thus,
the numerical procedure for calculation of $g_{v}^{(1)}$ can also
be straightforwardly applied to calculate $g_{v}^{(2)}$.

\begin{figure}[htbp]
\begin{centering}
\includegraphics[width=0.6\textwidth]{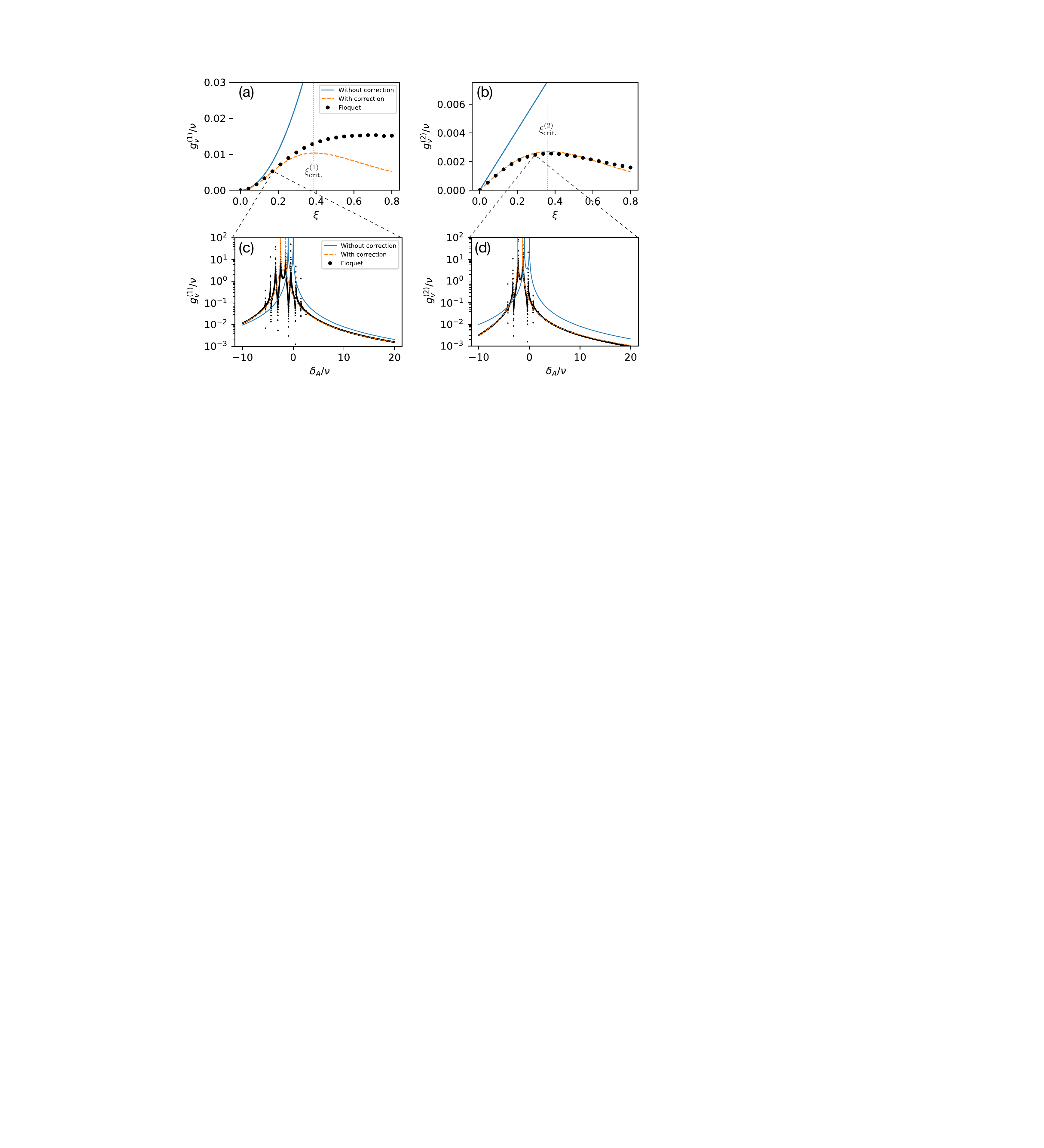}
\par\end{centering}
\centering{}\caption{Comparison of the coupling rate expressions with numerical Floquet
calculations. (a), (b) Coupling rates $g_{v}^{(1,2)}$ plotted as
a function of drive strength. (c), (d) Coupling rates plotted as a
function of the phonon mode detuning $\delta_{A}$. The uncorrected
coupling rates exhibit two resonant peaks, at $\delta_{A}=0$ and
$\delta_{A}+\nu=\delta_{B}=0$, corresponding to resonant processes
where phononic excitations in modes $A$ or $B$ are converted to
transmon excitations. Because of the AC Stark shift, these peaks are
red-shifted in both the numerical Floquet calculation and the corrected
expressions. The additional resonant peaks in the numerical calculation
correspond to multiphoton resonances where phononic excitations are converted to transmon excitations by exchanging an integer number of photons between the two drive fields~\cite{zhang2019}.
It is important to carefully avoid these peaks in the experiments. Parameters for all plots match those in Fig.~3 of the main text: $g_k/2\pi=10\si{\mega\hertz},$
$\delta_{A}/2\pi=100\si{\mega\hertz}$, $\nu/2\pi=10\si{\mega\hertz}$,
$\Delta\nu=\nu/10$. In order to account for the AC Stark shift, we
also specify $\alpha/2\pi=150\si{\mega\hertz}$, and we take $\delta_{1}/2\pi=1\text{GHz}$
in the calculation of $g_{v}^{(1)}$. In (c), $\xi_{1,2}=0.17$, and
in (d) $\xi_{1}=0.27$. \label{fig:stark_comparison}}
\end{figure}

In Figs.~\ref{fig:stark_comparison}(a) and (b), we calculate $g_{v}^{(1,2)}$ numerically as described above, and we compare the results with the analytical expressions (\ref{eq:gv1_tilde})
and (\ref{eq:gv2_tilde}). Good agreement is observed for weak drives
($\xi\lesssim0.4$ for the parameters used in the plots). Discrepancies
emerge at stronger drives, but this is expected because the corrections
are obtained perturbatively. In Fig.~\ref{fig:stark_comparison} (c),
(d), the coupling rates are plotted as a function of $\delta_{A}$
to make apparent the importance of the AC Stark shift corrections.
Due to the Stark shift, the corrected expressions and numerics are
both red-shifted relative to the uncorrected expressions. Were the
corrections not included, this relative shift would result in a systematic
overestimation of the coupling rates for blue-detuned phonon modes.

The AC Stark shift is responsible for the interesting non-monotonic
behavior of expressions (\ref{eq:gv1_tilde}) and (\ref{eq:gv2_tilde})
with $\xi$. Intuitively, this behavior is explained by the fact that
the Stark shift causes the qubit to move \emph{away} from the phonon
modes in frequency space. This reduces the participation of the phonons
in the qubit mode, therby reducing the coupling rate. When optimizing
$g_v$ so as to minimize the \texttt{SWAP} or \texttt{CZ} infidelity,
the non-monotonicity effectively restricts the drive amplitudes to
the range $\xi\leq\xi_{\text{crit.}}$, where $\xi_{\text{crit}.}$
is the value of $\xi$ for which $g_{v}$ is maximal.
For example, in Fig.~3 of the main text, the virtual couplings rates are restricted to $|g_v^{(1)}|/2\pi < \SI{100}{\kilo\hertz}$ and $|g_v^{(2)}|/2\pi < \SI{25}{\kilo\hertz}$, in accordance with the maximal values of $g_v^{(1,2)}$ obtained in Fig.~\ref{fig:stark_comparison}.
Good agreement
between the analytics and numerics is observed for $\xi\lesssim\xi_{\text{crit.}}$,
validating the use of expressions (\ref{eq:gv1_tilde}) and (\ref{eq:gv2_tilde}). 

The numerical calculations in Fig.~\ref{fig:stark_comparison} (a)
suggest that $g_{v}^{(1)}$ could be further increased by allowing
$\xi>\xi_{\text{crit.}}$. To be conservative, however, we choose
not to exploit this possibility; strong drives can induce deliterious processes
such as the multiphoton resonances described in Ref.~\cite{zhang2019}. 

This comparison illustrates the importance of the corrections derived
above and confirms that the virtual coupling rates are well-described
by expressions (\ref{eq:gv1_tilde}) and (\ref{eq:gv2_tilde}) for
the drive strengths considered in this work. 


\section{Engineering nonuniform mode spacing}
\label{sec:nonuniform}
As discussed in the main text, nonuniform mode spacing is necessary in order to ensure that the resonance conditions are nondegenerate, i.e.~to ensure that a given pair or triple of modes can be selectively coupled. In this section, we first formalize the meaning of nonuniform, then present several schemes for engineering nonuniformity in BAW and SAW systems---those in which modes are otherwise approximately uniformly spaced.
For concreteness, we also provide example schematics for BAW and SAW devices with engineered nonuniformity. Note that phononic crystal resonators are not generally plagued by such degeneracies, since mode frequencies can be controlled by engineering the geometry of each individual phononic resonator. 

\begin{figure}[htbp]
\begin{centering}
\includegraphics[width=0.45\textwidth]{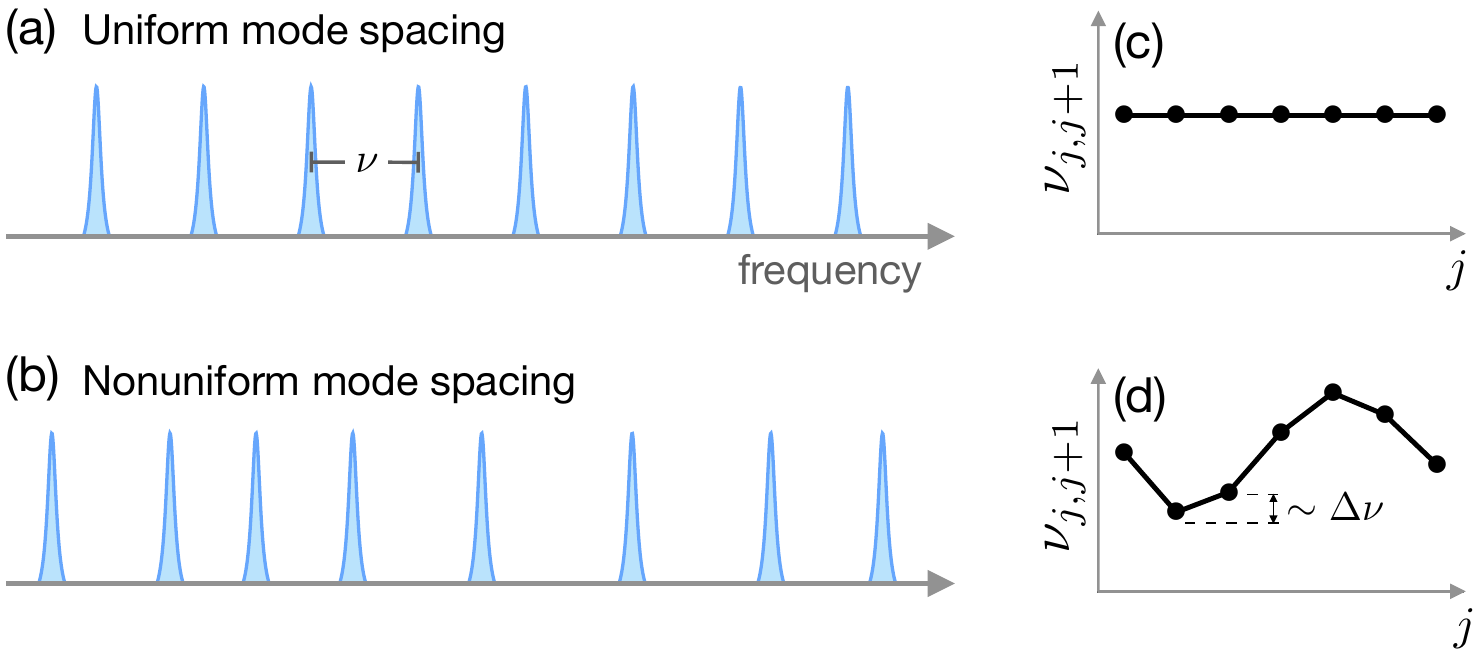}
\par\end{centering}
\centering{}\caption{
Sets of (a) uniformly and (b) nonuniformly spaced modes. 
(c,d) The frequency differences between successive modes are plotted to illustrate the behavior of $\nu_{j,j+1}$. (c) For uniformly spaced modes, $\nu_{j,j+1}$ is constant. (d) For nonuniformly spaced modes, $\nu_{j,j+1}$ varies on the scale of $\Delta\nu$.
 \label{fig:nonuniform_def}}
\end{figure}

As shown in Fig.~\ref{fig:nonuniform_def}, a set of modes is nonuniformly spaced if there exist mode pairs $\{i,j\}$ and $\{k,\ell\}$ for which $\nu_{ij} \neq \nu_{k\ell}$, where $\nu_{ij}=|\omega_i-\omega_j|$ is the frequency spacing between modes $i$ and $j$. 
In the context of multimode coupling, it is useful to quantify this nonuniformity as follows. 
Let $\mathcal{S}$ denote the set of all modes that are used to store quantum information, and let $\mathcal{P}$ denote the set of all mode pairs that one chooses to couple (we provide examples below). 
The connectivity of the system is then described by a graph with vertices $\mathcal{S}$ and edges $\mathcal{P}$. 
As a practically relevant measure of the nonuniformity, we define the quantity 
\begin{equation}
\Delta\nu = \min_{\{i,j\} \in \mathcal{P}} \left[\min_{\{k\in \mathcal{S},\ell\}\neq\{i,j\}}|\nu_{ij}-\nu_{k\ell}|\right],
\end{equation}
which lowerbounds the frequency selectivity of two-mode couplings. 
Explicitly, the beamsplitter resonance condition, $\omega_2-\omega_1 = \omega_B-\omega_A$ for a pair of modes $\{A,B\}\in \mathcal P$ is detuned from all other beamsplitter resonance conditions involving any mode in $\mathcal{S}$ by at least $\Delta\nu$. 
Highly selective virtual couplings thus require $g_v/\Delta\nu\ll 1$. Note that since $\Delta\nu$ depends on the choices of $\mathcal{S}$ and $\mathcal{P}$, there can exist a tradeoff between selectivity and the effective size and connectivity of the system.
The definition of $\Delta\nu$ can be straightforwardly generalized to the case of three-mode couplings.

Whether a given pair or triplet of modes can be selectively coupled depends on the structure of the nonuniformity, and 
in this regard it is convenient to classify different sorts of nonuniformity according to properties of $\nu_{j,j+1}$. We study two such classes in the examples below: \emph{point defect nonuniformities,} for which $\nu_{j,j+1}$ is constant except in the vicinity of a single defect, and \emph{periodic nonuniformities,} for which $\nu_{j,j+1}$ is periodic. Of course, other classes exist, but we focus on these two classes since instances can readily be engineered in cQAD systems.

\subsection{External mode hybridization}
\begin{figure}[htbp]
\begin{centering}
\includegraphics[width=0.7\textwidth]{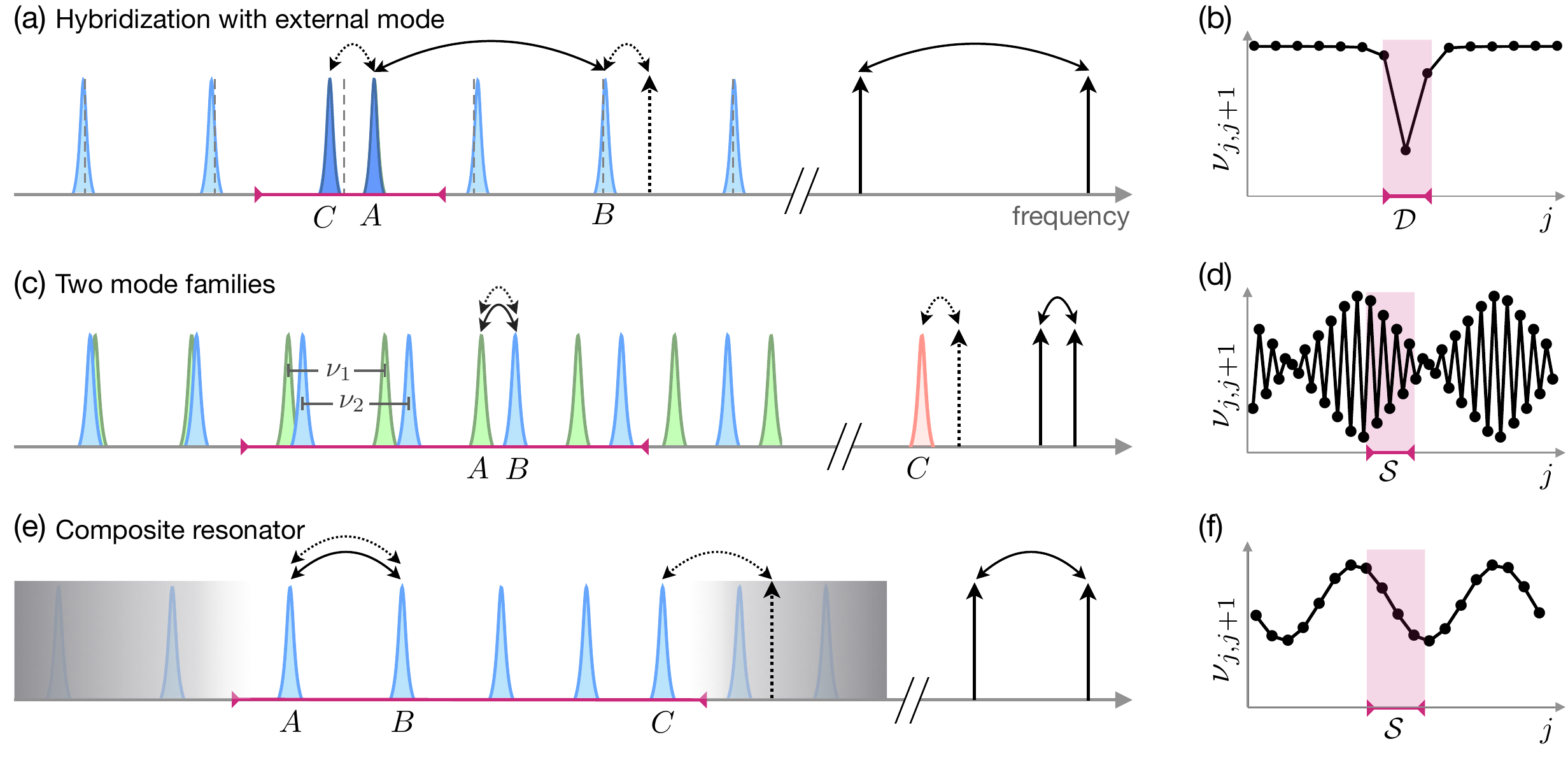}
\par\end{centering}
\centering{}\caption{Nonuniform mode spacing. 
(a) External mode hybridization. The coupling between phonons and an external mode causes strongly hybridized modes (dark blue) to deviate from the otherwise uniform spacing (dashed lines). The arrows show examples of how this nonuniformity gives rise to nondegenerate resonance conditions: modes $A$ and $B$ can be coupled by the applying drives indicated by solid arrows, while modes $A$, $B$, and $C$ can be coupled by applying the drive indicated by the dashed arrow.  
(b) Frequency differences shrink significantly within a bandwidth $\mathcal{D}$ of the external mode.  
(c) Two mode families. Simultaneously coupling the transmon to two mode families (blue, green) enables selective two-mode coupling between modes from different families. Selectivity is only guaranteed in a finite region $\mathcal{S}$, and an example of such a region is highlighted in (d). The use of an external mode $C$ enables selective three-mode coupling. 
(e) Composite resonator. Nonuniform mode spacing in composite resonators arises due to partial reflections at the interface(s). For example, with a single interface, a simple transfer matrix treatment~\cite{kharel2018b} reveals that the FSR is periodically modulated, as in (f). Selective three-mode coupling can be enabled by restricting the transmon phonon-coupling bandwidth (regions with negligible coupling are shaded in gray), or by using an external mode as in (c).
 \label{fig:nonuniform}}
\end{figure}

A point defect nonuniformity can be created by coupling the phonons to some external mode, such as a microwave resonator. 
As demonstrated in Ref.~\cite{han2016}, and sketched in Fig.~\ref{fig:nonuniform}(a,b), the resulting mode hybridization can significantly shift phonon mode frequencies within some bandwidth $\mathcal{D}$ of the external mode. 
The nonuniformity $\Delta\nu$ is dictated by the magnitude of these frequency shifts. For example, frequency shifts of order $1\si{\mega\hertz}$ were demonstrated in Ref.~\cite{han2016}.

This class of nonuniformity can enable selective coupling: selective two-mode coupling is possible if one or both involved modes lie in $\mathcal{D}$, and selective three-mode coupling is possible if two of the three involved modes lie in $\mathcal{D}$. Hence, the set $\mathcal{S}$ can include arbitrarily many modes, but the set $\mathcal{P}$ can only include mode pairs with at least one mode in $\mathcal{D}$. While modes outside of $\mathcal{D}$ cannot be directly coupled to one another, information from these modes can instead be swapped into modes in $\mathcal{D}$, manipulated, and swapped back. 
Note that the coherence of the external mode should be comparable to that of the phonons, lest the hybridization result in a significant increase in effective decay rates, and in general there may exist a tradeoff between increased nonuniformity and enhanced decay.

\subsection{Two phonon mode families}
\label{two_families}
Another approach is to create a periodic nonuniformity by simultaneously coupling the transmon to two families of phonon modes~\cite{kervinen2018} with different free spectral ranges (FSRs). 
While modes within each family are uniformly spaced, the FSR difference causes the spacing between modes from different families to vary, as shown in Fig.~\ref{fig:nonuniform}(c,d). 
This nonuniformity enables two modes from different families to be selectively coupled, but because of the periodicity, selectivity is only guaranteed over a finite bandwidth smaller than one period. 
With two mode families, a set $\mathcal{S}$ containing $\approx \nu/\Delta\nu$ modes can be found wherein any two modes from different families can be selectively coupled with $\Delta\nu =  |\nu_1-\nu_2|$, where $\nu_{1,2}$ are the FSRs of the two families.

By itself, the use of two mode families does not enable selective three-mode coupling\footnote{At least two modes from any three come from the same family, and since the modes in each family are uniformly spaced, there necessarily exists another set with the same resonance condition.}, but this limitation can be circumvented by coupling the transmon to one or more external modes. 
For example, the BAW devices of Refs~\cite{chu2017,chu2018} are housed in microwave cavities, and coupling the transmon to a high-Q cavity mode can enable selective three-mode coupling between the cavity and any pair of modes in $\mathcal{S}$. 
In a SAW device, the additional mode could come from another SAW resonator or a microwave resonator. The transmon itself could even serve as the external mode, but gate fidelities would then be directly limited by transmon coherence.
Ideally, the coherence of the external mode should be comparable to that of the phonons, lest it limit gate fidelity.

\subsection{Composite resonators}
Yet another approach is to employ a composite acoustic resonator, in which phonons propagate in media with different indices of refraction [Fig.~\ref{fig:nonuniform}(e,f)]. Reflections at the interfaces can give rise to a periodic modulation of the FSR~\cite{kharel2018b}. 
As in the case of two mode families, this periodic nonuniformity can enable selective two-mode coupling within a finite bandwidth $\mathcal{S}$, though the magnitudes of both $\mathcal{S}$ and $\Delta\nu$ depend on the nature of the modulation. 

Whether selective three-mode coupling within $\mathcal{S}$ is feasible depends on the of the specific nature of the FSR modulation. In cases where it is not already possible, selective three-mode coupling can be enabled by either coupling the transmon to some external mode, as previously described, or alternatively by restricting the bandwidth over which the transmon-phonon coupling is appreciable. 
For example, if the transmon-phonon coupling is only appreciable within $\mathcal{S}$, as in Fig.~\ref{fig:nonuniform}(e), then selective three-mode coupling is possible since the system contains an effectively finite number of nonuniformly spaced modes. 
In SAW systems, the coupling bandwidth can be tuned by changing the number of fingers in the interdigitated transducer\footnote{Because SAW resonators have finite bandwidth, care should be taken to avoid coupling to unconfined modes. This problem can be solved in general by engineering the transmon-phonon coupling bandwidth to lie within the SAW resonator bandwidth. The size of both bands can be tuned by varying the number of fingers in the respective interdigitated transducers~\cite{aref2016a,manenti2017}. }~\cite{aref2016a,manenti2017}. 
In BAW systems, the coupling bandwidth can be similarly tuned by changing the electromechanical transducer's geometry. 
For instance, in a transducer comprised of alternating layers of piezoelectric and non-piezoelectric materials, the spacing, thickness, and number of such layers could be chosen so that the coupling has a narrow response centered at a particular frequency, as in a Bragg reflector.

\subsection{Example schematics}
\begin{figure}[htbp]
\begin{centering}
\includegraphics[width=0.7\textwidth]{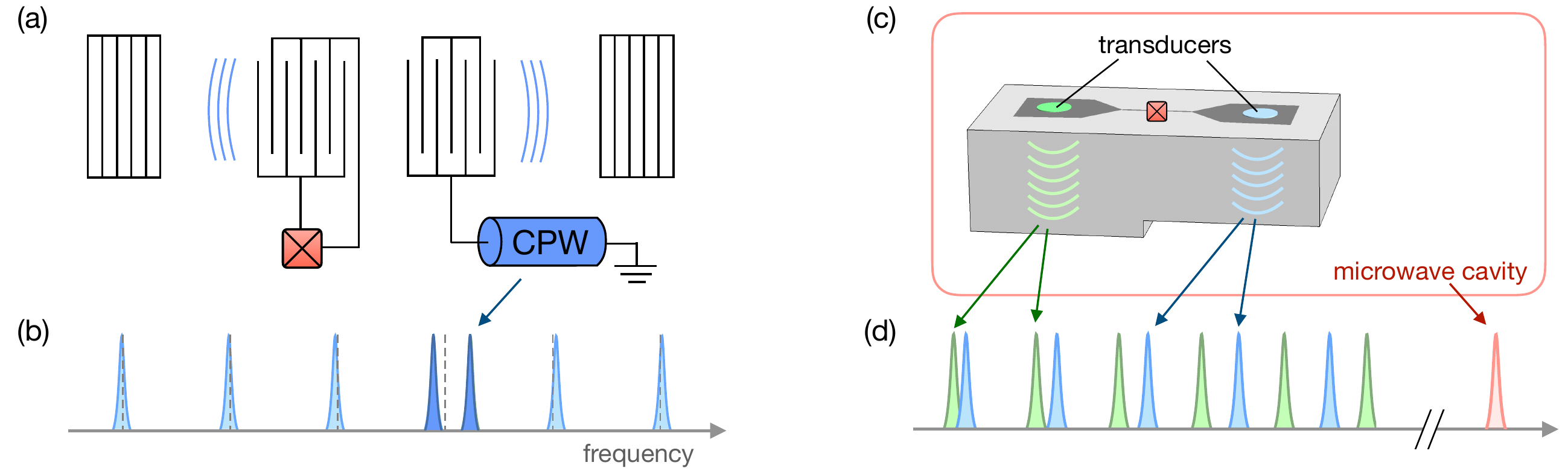}
\par\end{centering}
\centering{}\caption{SAW and BAW devices with engineered nonuniformity. (a) The modes of a SAW resonator are coupled to both a transmon and a coplanar waveguide (CPW) resonator. Hybridization with the resonator mode creates nonuniformity. (b) Mode frequencies of the device in (a). The CPW resonator mode and the phonon mode with which it most strongly hybridizes are shown in dark blue. (c) A 3D transmon couples to both a microwave cavity mode and to phonon modes from two BAW resonators with different FSRs (the difference is engineered by reducing the thickness of the substrate under one of the transducers). (d) Mode frequencies of the device in (c). 
 \label{fig:schematics}}
\end{figure}

For concreteness, in Fig.~\ref{fig:schematics} we provide example schematics for SAW and BAW devices in which nonuniformity is engineered according to the strategies described above. Fig.~\ref{fig:schematics}(a) shows a SAW device that exploits the external mode hybridization strategy. A SAW resonator is fabricated on a piezoelectric substrate, and coupling between the transmon and the phononic modes is enabled by an interdigitated capacitor. A superconducting coplanar waveguide resonator is also coupled to the phononic modes, and the hybridization of the phononic modes with the resonator mode creates the necessary nonuniformity.

Fig.~\ref{fig:schematics}(c) shows a BAW device that exploits the two mode families strategy. The device is based on those demonstrated in Refs.~\cite{chu2017,chu2018}; a three-dimensional (3D) transmon is housed inside a microwave cavity, and thin disks of piezoelectric material (transducers) fabricated in the transmon's pads enable the transmon to couple to BAW modes in the substrate. Two modifications have been made relative to the devices in Refs.~\cite{chu2017,chu2018}. First, an additional transducer has been added so that the transmon simultaneously couples to two families of modes. Second, the thickness of the substrate beneath one of the transducers has been reduced so that the two families have different FSRs. The microwave cavity mode, which dispersively couples to the transmon, provides the external mode necessary to enable selective three-mode couplings. We note that other elements, e.g.~a separate readout resonator for the transmon, can be integrated into 3D architectures in such a way that the transmon can be driven and measured without involving the cavity mode~\cite{axline2016}. 


\section{Detailed description of the cQAD QRAM}
In this section we provide a more complete description of the QRAM proposed in the main text. The operation of the quantum quantum routers is first discussed in detail, and then schemes for extracting data from either a classical or a quantum database are presented.

\subsection{Operation of a quantum router}

\begin{figure}[htbp]
\begin{center}
\includegraphics[width=0.9\textwidth]{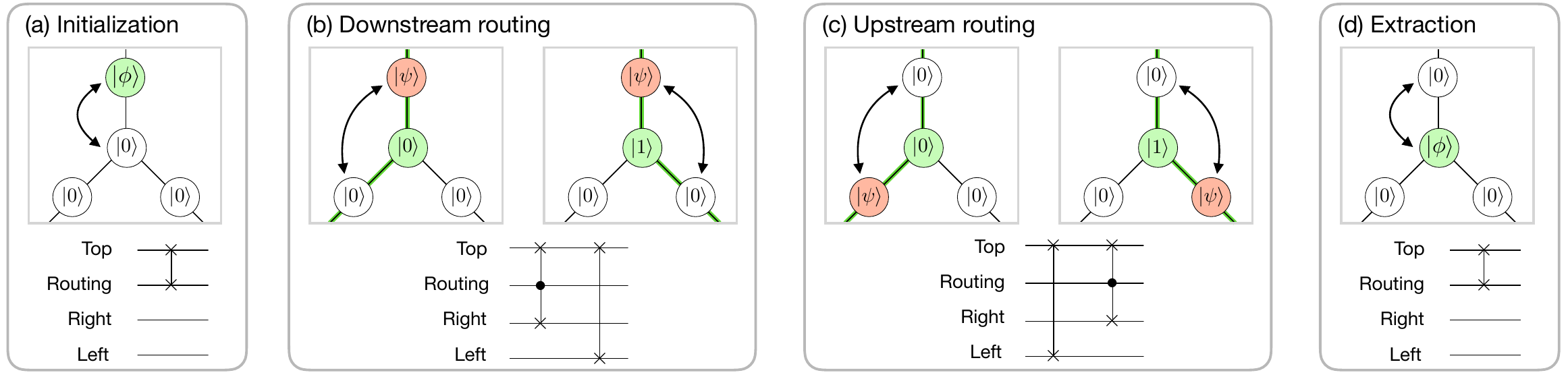}
\caption{Quantum router operation. (a) Initialization. An incoming address qubit $\ket{\phi}$ is swapped into the routing mode. (b) Downstream routing. An incoming qubit $\ket{\psi}$ is routed to the left(right) when the address qubit is in $\ket{0}$($\ket{1}$).  (c) Upstream routing. The router directs a qubit from either the right or left mode into the top mode conditioned on the state of the address qubit.  (d) Extraction. The address qubit is swapped into the top mode so that it can be routed back out of the tree. \label{fig:detailed_routing}  }
\end{center}
\end{figure}

During a QRAM query, the operation of each quantum router can be divided ino four stages: initialization, downstream routing, upstream routing, and extraction. In the initialization stage [Fig.~\ref{fig:detailed_routing}(a)], an incoming address qubit in the top mode is stored in the routing mode at the vertex so that it can control the routing of subsequent qubits. Initialization is performed simply by swapping the states of the top and routing modes. In the downstream routing stage [Fig.~\ref{fig:detailed_routing}(b)], the router directs incoming qubits to one of two output modes conditioned on the state of the address qubit. As shown in the corresponding circuit, this routing operation can be implemented using a \texttt{SWAP} gate and a controlled-\texttt{SWAP} gate. Note that the action of the controlled-\texttt{SWAP} gate is trivial when the routing qubit is in $\ket{0}$. Similarly,  the action of the \texttt{SWAP} gate is trivial when the routing qubit is in $\ket{1}$, since the affected modes are both in $\ket{0}$ after the controlled-\texttt{SWAP} has been performed.

Once all downstream routing is complete and the data has been transferred to the bus, the inverse operations (upstream routing and extraction) must be performed in order to disentangle the address and bus qubits from the routers.  Both the \texttt{SWAP} and controlled-\texttt{SWAP} unitaries are their own inverses, so the upstream routing [Fig.~\ref{fig:detailed_routing}(c)] and extraction [Fig.~\ref{fig:detailed_routing}(d)] operations can be implemented by time-reversing the downstream routing and initialization operations, respectively. 

\subsection{Database access schemes}

In a QRAM, the address and bus registers must necessarily be composed of qubits. However, the database itself can contain either classical or quantum data, depending on the intended application. How the database is accessed depends on the type of data. In the classical case, data can be directly copied into the bus, but in the quantum case the no-cloning theorem prevents one from copying quantum data. Instead, the bus can be entangled or swapped with the data, which generally leaves the address-bus system entangled with the database after a query. Additionally, while a QRAM should in general be able to both read from and write to the database, in many applications a quantum read-only memory (qROM) is sufficient~\cite{babbush2018,chakraborty2018,gilyen2018a}. This reduced functionality can be used to simplify the database access scheme. Below we present schemes for reading and writing both quantum and classical data, as well as a simplified scheme for read-only classical data.

\begin{figure}[htbp]
\begin{center}
\includegraphics[width=0.6\textwidth]{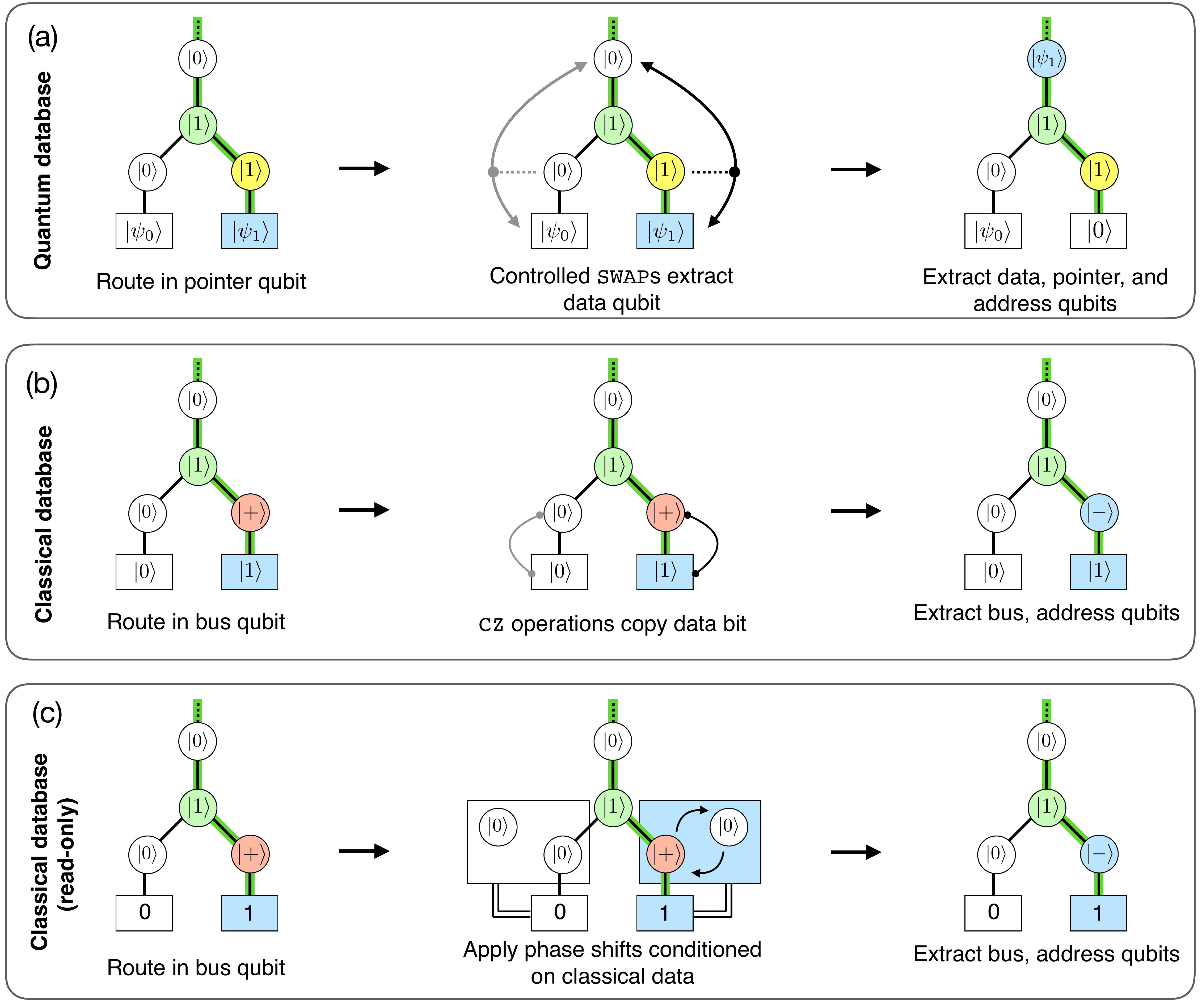}
\caption{Database access schemes for (a) quantum, (b) classical, and (c) classical read-only databases. 
(a) Quantum database access. A pointer qubit (yellow) is prepared in $\ket{1}$ and routed to the bottom of the tree, following the path carved by the address qubits  (green). Quantum data is stored in phononic modes, and the data is extracted via controlled-\texttt{SWAP} operations. The data, pointer, and address qubits are then sequentially routed out of the tree.  
(b) Classical database access. The bus qubit (red) is prepared in $\ket{+}$ and routed to the bottom of the tree. Classical data is encoded in the $\ket{0,1}$ Fock states of phononic modes, and the data is copied to the bus via \texttt{CZ} operations. The bus and address qubits are then sequentially routed out of the tree. 
(c) Classical read-only database access. Classical data is stored in a purely classical memory array and copied to the bus by applying corresponding phase shifts to each mode at the bottom of the tree.
\label{fig:database_access} }
\end{center}
\end{figure}

Figure~\ref{fig:database_access} illustrates procedures for accessing classical and quantum databases. 
To read quantum data from the database [Fig.~\ref{fig:database_access}(a)], we introduce a so-called pointer qubit, initially prepared in the state $\ket{1}$.
The purpose of this qubit is to indicate which database entries should be extracted. After all address qubits have been routed into place, the pointer qubit follows their path to the bottom of the tree. 
Modes at the bottom of the tree then serve as controls in controlled-\texttt{SWAP} operations that extract the desired data qubit. 
In each branch of the superposition, only the database entry adjacent to the pointer qubit is swapped out. 
The data, pointer, and address qubits are then routed out of the tree in sequence to complete the query. To write quantum data to the database\footnote{Writing either classical or quantum data to the database generally leaves the database in a superposition, unless entries are written one by one.}, the entire procedure can be run in reverse. 
Note that, if a query is error-free, the pointer qubit remains in the state $\ket{1}$ throughout and is disentangled from the rest of the system at the end of the query. 
Measuring this qubit after the query could thus provide a useful means of detecting errors.  

In the classical case, a data bit $D_j$ can be encoded in a mode as the Fock state $\ket{D_j}$. The operations described above for the quantum case then suffice to read or write classical data. 
However, since the read operation involves swapping the bus and data qubits, data is removed from the database during each query. 
In the classical case, data can instead be \emph{copied} to the bus [Fig.~\ref{fig:database_access}(b)], which ensures that queries leave the database undisturbed. 
To copy data, the bus qubit is initially prepared in the state $(\ket{0}+\ket{1})/\sqrt{2}\equiv \ket{+}$, and routed to the bottom of the tree.
The desired classical bit is then copied to the bus by performing a \texttt{CZ} operation between the data and bus qubits; if $\ket{D_j}=\ket{0}$, the bus is unaffected, but if $\ket{D_j}=\ket{1}$, the \texttt{CZ} operation flips $\ket{+}$ to $\ket{-}$, hence encoding the bit in the $\ket{\pm}$ basis. Importantly, this operation does not disturb other modes, since these phase shifts leave the state $\ket{0}$ unaffected. 
The bus and address qubits are then routed out of the tree to complete the query.

In the classical read-only case [Fig.~\ref{fig:database_access}(c)], it is not necessary to encode data in the Fock states of phonon modes.
Instead, the entire database can be stored in a purely classical memory array so that it is not susceptible to decoherence. 
To query such a database, the bus is again prepared in $\ket{+}$ and routed to the bottom of the tree. The desired classical bit is copied to the bus by applying phase shifts to each mode at the bottom of the tree in accordance with the database entries.
If $D_j=1$, a phase shift of $-1$ is applied to the $j^{th}$ mode, and if $D_j=0$, no phase shift is applied. 
As discussed in the main text, phase shifts can be imparted to a mode by tuning the relative driving phases during a consecutive sequence of two \texttt{SWAP} operations with an ancillary mode. 


\section{Derivation of gate fidelity expressions}
\label{gate_fidelities}

Given a quantum process $\mathcal E$ constituting a noisy implementation of a gate $U$, the gate fidelity is typically defined as 
\begin{equation}
    \mathcal F = \int d\psi \braket{\psi|U^\dagger \mathcal E(\psi) U |\psi},
\end{equation}
where the integral is over the uniform measure $d\psi$ on state space, with normalization $\int d\psi = 1$.  For an $n$-qubit gate, the states $\ket{\psi}$ are elements of the Hilbert space of the $n$ qubits involved in the gate. With this definition, crosstalk with modes not involved in the gate is quantified \emph{implicitly} via its deleterious effects on qubits involved in the gate, and the fidelity is independent of whether other modes are used to store quantum information.

Other definitions are possible in the context of multimode systems. For example, given a multimode system in which $M$ modes are used to store qubits, one can consider an alternate definition of the fidelity,
\begin{equation}
    \tilde{\mathcal F} = \int d\psi \braket{\psi|\tilde{U}^\dagger \tilde{\mathcal E}(\psi) \tilde{U} |\psi},
    \label{eq:alt_fid_def}
\end{equation}
where $\tilde{U} = U \otimes \mathbb{1}^{\otimes(M-n)}$ is the tensor product of the ideal $n$-qubit gate and the identity operation on the remaining $M-n$ qubits, $\tilde{\mathcal{E}}$ is the corresponding noisy implementation of the gate, and the states $\ket{\psi}$ are elements of the Hilbert space of all $M$ qubits in the system. With this definition, crosstalk is accounted for \emph{explicitly} via deviations of $\tilde{ \mathcal E}$ from the ideal operation $\tilde U$, and the fidelity generally depends on $M$.

In this section, we compute the direct and virtual gate fidelities according to both of the metrics, $\mathcal{F}$ and $\tilde{\mathcal{F}}$, which we refer to as the \emph{local} and \emph{global} fidelities, respectively.
The local fidelity $\mathcal{F}$ is commonly used in the literature in the context of multi qubit architectures~\cite{barends2014,wright2019} and has the advantage that it is largely independent of implementation differences between BAW, SAW, and phononic crystal devices. 
In contrast, the global fidelity $\tilde{\mathcal{F}}$ is a more comprehensive metric, but it depends sensitively on implementation details, making it challenging to analyze in general. 
In the main text we thus compute the local fidelity so as to facilitate a straightforward and implementation-independent comparison with other architectures.
For concreteness, though, we compute the global infidelity for a particular implementation based on a BAW device below, and we show that the main claim---that virtual gates can offer higher fidelities for long-lived phonon modes---holds when using either metric. A closely related discussion can be found in Ref.~\cite{pechal2018}.

\subsection{Local infidelity}
We now derive the expressions for the local infidelity $1-\mathcal{F}$ given in the main text (Eqs.~3 and 4).
There are two dominant contributions to the infidelity: decoherence and spectral crowding (crosstalk). The contributions to the direct and virtual gate infidelities from decoherence are, respectively,
\begin{equation*}
     (\kappa+\gamma)t_d \quad \text{and} \quad  \bar{\kappa}_\gamma t_v,
\end{equation*}
where $t_d = (c_d \pi/2 g_d)$ and $t_v = (c_v \pi/2 g_v)$ are the total gate times, and $\kappa+\gamma$ and $\bar{\kappa}_\gamma$ are the total decoherence rates for the direct and virtual gates respectively. For two-mode couplings, the total decoherence rate is $\bar{\kappa}_\gamma = \kappa_\gamma^A + \kappa_\gamma^B$, while for three-mode couplings the rate is $\bar{\kappa}_\gamma =( \kappa_\gamma^A + \kappa_\gamma^B + \kappa_\gamma^C)/2$. The factor of $1/2$ in the latter expression results from averaging over the gate duration; when there is a phonon each in modes $A$ and $B$, the rate is $\kappa_\gamma^A + \kappa_\gamma^B$, but once these two phonons have been converted into a single phonon in mode $C$ the rate is $\kappa_\gamma^C$.  
Note that we have neglected the inverse Purcell enhancement in the direct case under the assumption that this enhancement is negligible relative to $\gamma$.

The presence of other modes also contributes to the infidelity, regardless of whether these other modes are used to store quantum information. When performing a gate, transitions between the modes involved in the gate and the other modes are driven off-resonantly, and we approximate the spectral crowding infidelity as the probability that one of these unwanted transitions occurs. This probability is computed in Ref.~\cite{pechal2018} for direct gates: assuming a set of uniformly spaced modes with free spectral range $\nu$, the infidelity is approximately\footnote{Following Ref.~\cite{pechal2018}, we neglect a constant prefactor of order 1 on the right hand side, with the justification that $(g_d/\nu)^2$ is actually a pessimistic upper bound; the spectral crowding infidelity can be reduced by smoothly ramping up the drives.}
\begin{equation*}
    \sum_{n}\left(\frac{g_{d}}{\delta_n}\right)^{2}\approx \left(\frac{g_d}{\nu}\right)^{2},
\end{equation*}
where the sum on the left runs over all unwanted transmon-mode transitions, each detuned by successive multiples of $\nu$, i.e.~$\delta_n = \{\pm \nu,\pm 2\nu ,\ldots \}$. Importantly, this infidelity is independent of the total number of modes in the system. Similarly, for virtual gates the spectral crowding infidelity is approximately
\begin{equation*}
    \sum_{n}\left(\frac{g_{v}}{\delta_n'}\right)^{2}\approx \left(\frac{g_v}{\Delta\nu}\right)^{2},
\end{equation*}
where the sum runs over all unwanted virtual couplings that affect modes involved in the gate. More precisely, $n$ indexes all unwanted two-mode (three-mode) couplings which involve at least one mode from the set $\{A,B\}$ ($\{A,B,C\}$), and $\delta_n'$ is the resonance condition detuning between the $n^{th}$ unwanted coupling and the desired coupling. For example, in the two-mode case, these detunings are of the form $\delta'_n = (\omega_B-\omega_A) - (\omega_i - \omega_j)$, with either $i$ or $j \in \{A,B\}$. In performing the sum we have assumed that unwanted couplings are detuned by successive multiples of $\Delta\nu$, i.e.~$\delta_n' = \{\pm \Delta\nu, \pm 2\Delta\nu ,\ldots \}$. In general the $\delta_n'$ depend on the specific structure of the nonuniformity, but the scaling $(g_v/\Delta\nu)^2$ holds regardless.

Summing the contributions from decoherence and spectral crowding gives the infidelity approximations used in the main text,
\begin{align}
1-\mathcal{{F}}_{d}&\approx (\gamma+\kappa_\gamma)\left[\frac{c_{d}\pi}{2g_{d}}\right]+\left(\frac{g_{d}}{\nu}\right)^{2},\label{eq:F_d_local}\\
1-\mathcal{{F}}_{v}&\approx  \bar{\kappa}_\gamma \left[\frac{c_{v}\pi}{2g_{v}}\right]+\left(\frac{g_{v}}{\Delta\nu}\right)^{2}.\label{eq:F_v_local}
\end{align}
Note that these first-order expressions are valid only in the limit $1-\mathcal{F}\ll 1$. When plotting the infidelity, we include higher-order terms to ensure the output is restricted to [0,1]; the above expressions are of the form $k t + \varepsilon$, but we plot corresponding higher-order expressions $1- e^{- k t}(1-\varepsilon)$.

\subsection{Global infidelity}
We now compute the global infidelities $1-\tilde{\mathcal{F}}$ for direct and virtual gates. As before, decoherence and spectral crowding are the two dominant contributions to the infidelity. For a multimode system in which $M\geq2$ modes are used to store qubits, the total decoherence contribution to the direct gate infidelity $1-\tilde{\mathcal{F}}_d$ is
\begin{equation*}
    \left[ (\kappa+\gamma) +(M-2)\kappa\right]t_d,
\end{equation*}
since the two qubits involved in the gate experience a total decoherence rate of approximately $\kappa + \gamma$, while the remaining $M-2$ qubits are left idling in their respective resonator modes during the gate. 
The total decoherence contribution to the virtual gate infidelity $1-\tilde{\mathcal{F}}_v$ is $\bar{\kappa}_\gamma t_v$, where $\bar{\kappa}_\gamma$ is the total decoherence rate, which now includes contributions from all $M$ modes,
\begin{equation*}
\bar{\kappa}_\gamma = 
    \begin{cases} 
       M\kappa_\gamma & \text{for two-mode coupling,}\\
       \left(M-\frac{1}{2}\right)\kappa_\gamma & \text{for three-mode coupling.}
   \end{cases}
\end{equation*}
The three-mode coupling expression contains $(M-1/2)$ as opposed to $M$ because the conversion of two phonons into a single phonon and back reduces the average decoherence rate.

Next, consider the contribution from spectral crowding. 
Both direct and virtual gates require frequency selectivity, and spectral crowding infidelity results from the limited ability to resolve a desired transition from others which are nearby in frequency space. In order to compute the \emph{global} spectral crowding infidelity, it is thus necessary to specify the full distribution of transition frequencies. Because these distributions can vary from platform to platform, however, it is challenging to make general statements about the spectral crowding infidelity. Computing this infidelity for the multitude of possible implementations is beyond the scope of this work, so in the following we consider a particular implementation based on a BAW or SAW device as an illustrative example. We stress that the following discussion applies only to this particular implementation. 

Consider a multimode cQAD device in which the acoustic modes are supported in a BAW or SAW resonator. In the context of direct gates, we consider a single multimode resonator, in which modes are approximately uniformly spaced with FSR $\nu$. In the context of virtual gates, nonuniform mode spacing is required, and we consider the case where this nonuniformity is engineered through the use of two mode families with different FSRs, $|\nu_1-\nu_2|=\Delta \nu \neq 0$, as discussed in Sec.~\ref{two_families}.  

The spectral crowding contributions to the direct and virtual infidelities, $1-\tilde{\mathcal{F}}_{d}$ and $1-\tilde{\mathcal{F}}_v$, can be respectively written as
\begin{equation*}
    S_d\left(\frac{g_d}{\nu}\right)^2 \quad\text{and}\quad S_v\left(\frac{g_v}{\Delta\nu}\right)^2,
\end{equation*}
where the coefficients $S_{d,v}$ remain to be determined. The spectral crowding contribution to $1-\tilde{\mathcal{F}}_d$ is given by
\begin{equation*}
     \sum_{n}\left(\frac{g_{d}}{\delta_n}\right)^{2} = \frac{\pi^2}{3} \left(\frac{g_d}{\nu}\right)^{2},
\end{equation*}
where $\delta_n = \{\pm \nu,\pm 2\nu ,\ldots \}$ as before. Thus, $S_d = \pi^2/3$.
Notice that unwanted transmon-phonon transitions contribute to the direct gate infidelity regardless of whether qubits are encoded in the other modes, cf.~the spectral crowding term in $1-\mathcal{F}_d$.

\begin{figure}[htbp]
\begin{center}
\includegraphics[width=\textwidth]{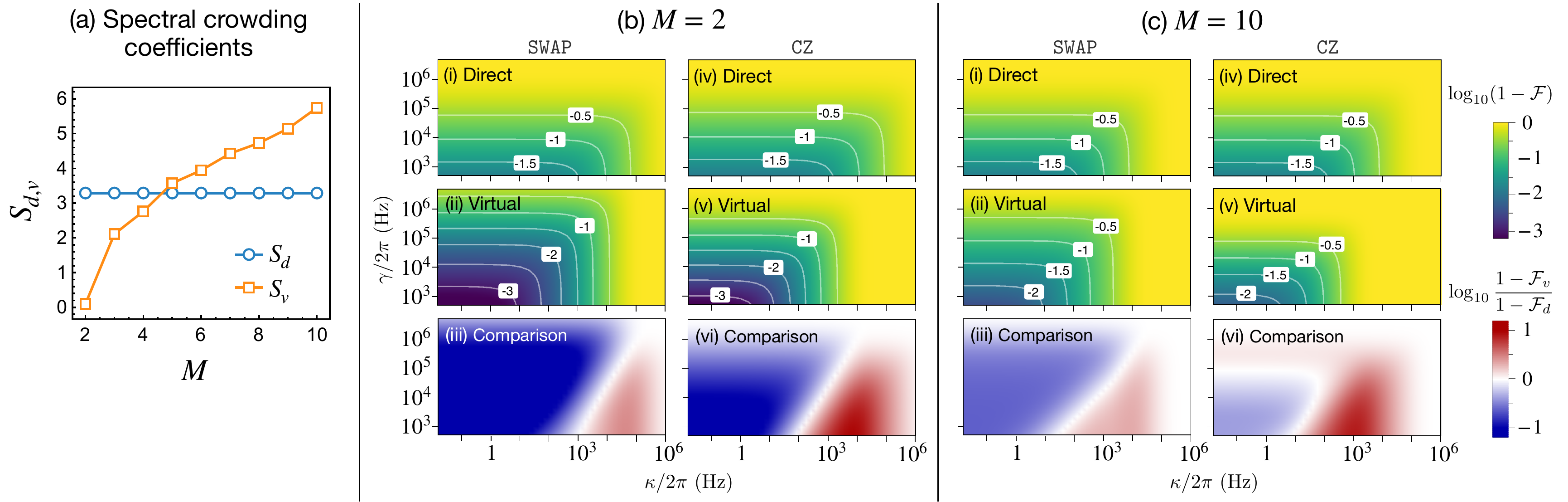}
\caption{Global infidelities. (a) Spectral crowding coefficients $S_{d,v}$. (b,c) Global infidelities $1-\tilde{\mathcal{F}}_{d,v}$ for (b) $M=2$ and (c) $M=10$.   Plot parameters are the same as in Fig.~3 of the main text, with the exception of $\Delta\nu$, which has been reduced to $\Delta\nu/2\pi =0.85$ MHz so that the number of selectively addressable modes ($\sim\nu/\Delta\nu$) is $ >10$.  
\label{fig:global} }
\end{center}
\end{figure}

We compute the spectral crowding contribution to $1-\tilde{\mathcal{F}}_v$ numerically by generating a large array of modes and directly computing the sum $\sum_k'(g_v/\delta_n')^2$. Here, $n$ indexes all unwanted virtual couplings that affect at least one of the $M$ qubits (cf.~the local infidelity, where only unwanted couplings that affected at least one of the qubits \emph{involved in the gate} were considered).
We take the average over all possible desired transitions among the $M$ modes to obtain $S_v$.  As shown in Fig.~\ref{fig:global}(a), we find that that $S_v$ increases with $M$ for fixed $\nu/\Delta\nu$. Note that for the particular implementation discussed in Sec.~\ref{two_families}, $S_v$ is the same for both engineered two- and three-mode couplings.

Summing the contributions from decoherence and spectral crowding gives
\begin{align}
1-\tilde{\mathcal{{F}}}_{d}&\approx  
    \left[ (\kappa+\gamma) +(M-2)\kappa\right]t_d
    + S_d\left(\frac{g_d}{\nu}\right)^2
    ,\label{eq:F_d_global}\\
1-\tilde{\mathcal{{F}}}_{v}&\approx 
     M\bar{\kappa}_\gamma t_v
    +S_v\left(\frac{g_v}{\Delta\nu}\right)^2
.\label{eq:F_v_global}
\end{align}
As before, $g_{d,v}$ may be tuned to minimize the infidelities, and we plot the minimal infidelities for the cases of $M=2$ and $M=10$ in Fig~\ref{fig:global}(b,c). 
For the $M=2$ case, the plots look very similar to the local infidelity plots in Fig.~3 of the main text; the only difference results from the fact that we keep track of the prefactors $S_d$ and $S_v$ in the global infidelity. For the $M=10$ case, reductions in both direct and virtual gate fidelities are observed relative to the $M=2$ case, as expected. The reductions are more pronounced in the virtual case, due in part to the fact that $S_v$ scales with $M$ while $S_d$ does not.   
This scaling difference is a property of the specific BAW/SAW implementation that we consider here, and in fact we have chosen this example to illustrate that such differences can exist. Because of this scaling difference, this example effectively constitutes a worst-case scenario for the comparison of direct and virtual gate fidelities.  
Despite this, there are still large regions of parameter space where virtual gates offer higher fidelities.

Two additional remarks are necessary. First, we stress that the spectral crowding infidelities can vary from platform to platform, so the above results are specific to the BAW/SAW implementation that we have considered. In phononic crystal devices, for example, $S_{d,v}$ will have different a dependence on $M$. In these devices the mode frequencies must be distributed within a fixed bandgap, so the effective FSR scales with $1/M$, and hence $S_d$ scales with $M^2$~\cite{pechal2018}. It is possible to arrange the mode frequencies so that $S_v$ also scales with $M^2$. As a result, the comparison between direct and virtual gates would be less affected by the choice of $M$ in such devices. 

Second, the above analysis illustrates that for large $M$, it will be impractical to control all $M$ qubits using only a single transmon. A more scalable approach is to construct larger devices out of smaller modules, where each module contains its own acoustic modes and transmon. In this context, the optimal number of modes \emph{per module} will typically to be closer to 10, rather than to 100 or 1000.  
For example, the performance of a module can be quantified using the quantum volume~\cite{bishop2017,pechal2018},
\begin{equation}
    V = \max_M \left[\min(M, d(M)) \right]^2,
\end{equation}
where $d(M) = 1/M(1-\tilde{\mathcal{F}})$ is the maximal circuit depth\footnote{This assumes no parallelization. Unlike direct gates, virtual gates can be performed in parallel, and parallel operation can increase the quantum volume.}. For the BAW/SAW device discussed above, we find that the $M$ which optimizes the quantum volume satisfies $M\leq12$ across the entire parameter space shown in the plots of Fig.~\ref{fig:global}(b,c).
Similarly, in Ref.~\cite{pechal2018}, $M\approx15$ was found to optimize the quantum volume of multimode cQAD devices based on phononic crystal resonators, assuming realistic future device parameters.


\section{Effects of phonon mode frequency shifts}
\label{frequency_shifts}
Due to hybridization with the transmon, the acoustic modes in multimode cQAD systems can experience frequency shifts caused by transmon drives (AC Stark shifts), or the presence of phonons (Kerr interactions). In this section, we discuss the effects of such shifts on the two- and three-mode resonance conditions given in the main text. 

\subsection{AC Stark shifts}
The AC Stark shift terms (Table I, Row 2),
\begin{equation}
    -2\alpha\sum_{k}|\xi_{k}|^{2}|\lambda_{j}|^{2}m_{j}^{\dagger}m_{j}\equiv  S_{j}m_{j}^{\dagger}m_{j},
\end{equation}
describe a drive-dependent shift of the phonon mode frequencies. In order to engineer resonant interactions in the presence of these shifts, the drive strengths and frequencies must be carefully chosen so that interactions between the \emph{Stark-shifted modes} are brought on resonance~\cite{zhang2019}. For example, a resonant beamsplitter-type coupling between modes $A$ and $B$ can be engineered by applying two drive tones tones whose frequencies $\omega_{1,2}$ satisfy,
\begin{equation}
  \omega_{2}-\omega_{1}=(\omega_{B}+S_{B})-(\omega_{A}+S_{A}).  
\end{equation}
Since the shifts $S_j$ are calculable, it is straightforward to ensure that this modified resonance condition is satisfied. For example, one can fix the drive amplitudes (which, when the drives are far-detuned, approximately fixes the $S_j$), then tune the frequencies to satisfy the modified condition.

As discussed in the main text, nonuniform mode spacing is necessary to ensure that couplings between modes can be selectively engineered. It is crucial that this nonuniformity persists in the presence of the phonon modes' Stark shifts, lest the resonance conditions become degenerate. In general, resonance conditions will remain nondegenerate provided that the Stark shifts are small relative to the nonuniformity, $S\ll\Delta\nu$. Since the Stark shifts are comparable in magnitude to the engineered two-mode couplings, $S\sim g_{v}^{(1)}$, the same degree of nonuniformity needed to guarantee highly selective couplings, $\Delta\nu\gg g_{v}^{(1)}$, also suffices to ensure that the Stark shifts do not create additional degeneracies in the resonance conditions. 

\subsection{Self- and Cross-Kerr interactions}
We next consider the effects of and self- and cross- Kerr interactions (Table I, row 5),
\begin{align}
    -\frac{\alpha}{2}|\lambda_{j}|^{4}m_{j}^{\dagger}m_{j}^{\dagger}m_{j}m_{j} 
    &\equiv -\frac{\chi_{jj}}{2}m_{j}^{\dagger}m_{j}^{\dagger} m_{j}m_{j}, \\
    \text{and}\quad   
    -2\alpha|\lambda_{j}|^{2}|\lambda_{k}|^{2}m_{j}^{\dagger}m_{k}^{\dagger}m_{j}m_{k} 
    &\equiv -\chi_{jk}m_{j}^{\dagger}m_{j}m_{k}^{\dagger}m_{k}, \quad (j\neq k).
\end{align}
Below, we discuss the effects of these interactions on engineered two-mode and three-mode couplings.
Unlike the AC Stark shifts, the effects of these terms on the resonance conditions cannot be compensated in general. 
However, we show that the infidelity associated with these terms can be made negligible relative to the infidelity associated with decoherence and spectral crowding. 

\subsubsection{Effects of Kerr on two-mode couplings}
For an engineered two-mode interaction $H^{(1)}= (g_v^{(1)} m_A^\dagger m_B + \text{H.c.})$, we first consider the effects of cross- and self-Kerr terms involving only modes $A$ and $B$. Recall that we consider encodings with $\leq1$ phonon per mode, so only the initial state $\ket{1,1}_{A,B}$ is affected by these interactions.  From this initial state, the coupling $H^{(1)}$ induces evolution within the subspace $\{\ket{11},\ket{20},\ket{02} \}$. In this subspace, the full Hamiltonian $H = (H^{(1)} + \text{Kerr terms})$ is
\begin{equation}
    H =
    \begin{pmatrix}
    -\chi_{AB} & \sqrt 2 g_v^{(1)} & \sqrt 2 g_v^{(1)} \\
    \sqrt 2 g_v^{*(1)} & -\chi_{AA} & 0 \\
    \sqrt 2 g_v^{*(1)} & 0 & -\chi_{BB} 
    \end{pmatrix}.
\end{equation}
We define the infidelity in terms of the overlap between the final state obtained by evolution under this Hamiltonian, $\exp(-i H t)\ket{11}$, and the ideal final state in the absence of these Kerr terms, $\exp(-i H^{(1)} t)\ket{11}$,
\begin{equation}
    1-\mathcal{F} = 1-|\braket{11|e^{i H^{(1)} t}e^{-i H t}|11}|^2.
\end{equation}
The infidelity can be approximated by making two assumptions: (i) that $\chi \ll g_v^{(1)}$, and (ii) that $\chi_{AA}\approx \chi_{BB} \approx\chi_{AB}/2$. The first assumption is justified by the fact that the transmon-induced self and cross-Kerr interactions, which are fourth order in $\lambda = g_j/\delta_j \ll1$, are typically much weaker than the engineered coupling $g_v^{(1)}$, which is only second order in $\lambda$. For example, in Ref.~\cite{gao2018}, $\chi_{AB}/2\pi\approx 1$ kHz, while $g_v^{(1)} / 2\pi \approx 30$ kHz. The second assumption is justified in the regime where $\lambda_A \approx \lambda_B$, i.e.~when the transmon-phonon hybridization is comparable for modes $A$ and $B$. This will be the case, for example, when the bare couplings are comparable, $g_A \approx g_B$, and the relative difference in the detunings is small, $|\delta_{B}-\delta_{A}|/\delta_{A,B}\ll 1$.
Under these assumptions, one finds that the infidelity scales quadratically with the small parameter $(\chi_{AB}/g_v^{(1)})$,
\begin{equation}
1-\mathcal{F} \approx \frac{1}{64}\sin^4\left(2 g_v^{(1)} t\right)\left(\chi_{AB}/g_v^{(1)}\right)^2 + O\left[\left(\chi_{AB}/g_v^{(1)}\right)^4\right].
\label{eq:Kerr_infid1}
\end{equation}
We provide numerical estimates of this infidelity below. 

In multimode systems, cross-Kerr interactions with other modes $j\notin\{A,B\}$ can also contribute to the infidelity. The effect of such interactions is to modify the two-mode resonance condition,
\begin{equation}
    \omega_{2}-\omega_{1}=\omega_{B}-\omega_{A}+ D^{(1)},
\end{equation}
where $D^{(1)}=\sum_{j}(\chi_{Aj}-\chi_{Bj})n_{j}$, and $n_{j}\in\{0,1\}$ denotes the number of phonons in mode $j$. Since the $n_{j}$ are not known \textit{a priori}, such shifts cannot generally be compensated and contribute to the gate infidelity. When the detuning of the drives from the proper resonance condition is small, $D^{(1)}\ll g_v^{(1)}$, the associated contribution to the infidelity is approximately
\begin{equation}
1-\mathcal{F}\approx (D^{(1)}/g_v^{(1)})^2.
\label{eq:Kerr_infid2}
\end{equation}

To numerically estimate the infidelities typically associated with these cross- and self-Kerr interactions, we consider the experiment performed in Ref.~\cite{gao2018}, in which a beamsplitter-type coupling was engineered between two microwave cavity modes via their mutual coupling to a transmon\footnote{Though the experiment used electromagnetic modes, rather than acoustic modes, the engineered coupling and transmon-induced Kerr interactions are expected to have similar magnitudes since they arise via the same mechanisms.}. In that work, engineered couplings as large as $g_v^{(1)}/2\pi = 30$kHz were demonstrated, and Kerr interactions were measured to be $\chi/2\pi \lesssim 1$kHz.  With these parameters, the infidelity contribution~(\ref{eq:Kerr_infid1}) is $< 10^{-4}$.

To estimate the contribution (\ref{eq:Kerr_infid2}), we note that the cross-Kerr differences $(\chi_{Aj}-\chi_{Bj})$ will typically be much smaller than the cross-Kerr interactions themselves. 
For example, when the frequency separation between the modes satisfies $\nu \ll \delta_{A,B}$, 
and assuming $g_A\approx g_B$, 
\begin{equation}
    (\chi_{Aj}-\chi_{Bj})\approx 2 \frac{\nu}{ \delta_{A}} \chi_{Aj} \ll \chi_{Aj}.
\end{equation}
For the parameters in Fig.~3 of the main text, typical cross-Kerr differences are $\sim100$ Hz .
The contribution~(\ref{eq:Kerr_infid2}) also depends on the total number of phonons, which will usually be order $10$ or fewer (see Sec.~\ref{gate_fidelities}). Thus, typically $D^{(1)}/2\pi \approx 100 \text{Hz} - 1$ kHz, and the infidelity contribution~(\ref{eq:Kerr_infid2}) is $10^{-5} - 10^{-3}$. 
We conclude that, for two-mode couplings, the infidelity associated with self- and cross-Kerr interactions is typically negligible relative to the infidelity associated with decoherence and spectral crowding.

\subsubsection{Effects of Kerr on three-mode couplings}
Next we study the effects of Kerr interactions on engineered three-mode couplings. Given an engineered three-mode interaction $H^{(2)}= (g_v^{(2)} m_A m_B m_C^\dagger + \text{H.c.})$, we first consider the effects of cross- and self-Kerr terms involving only modes $A$, $B$, and $C$.  As discussed in the main text, the initial states we consider have the form $\ket{n_{A}\leq1,n_{B}\leq1,n_{C}=0}$, and of these states only $\ket{1,1,0}$ is affected nontrivially by such Kerr interactions. This is also the only initial state on which the engineered coupling $H^{(2)}$ acts nontrivially. Thus, the effects of cross-Kerr and self-Kerr interactions between the modes $A$, $B$, and $C$ can be fully compensated simply by replacing $(\omega_A +\omega_B)\rightarrow (\omega_A +\omega_B-\chi_{AB})$ in the resonance condition. 

Cross-Kerr interactions with other modes $j\notin\{A,B,C\}$ yield the modified resonance condition,
\begin{equation}
    \omega_{1}=\omega_{A}+\omega_{B}-\omega_{C}+ D^{(2)}
\end{equation}
where $D^{(2)}=\sum_{j}(\chi_{Cj}-\chi_{Aj}-\chi_{Bj})n_{j}$. Since $n_{j}$ is not known \textit{a priori} these shifts cannot generally be compensated, and the associated contribution to the infidelity would be approximately $ (g_v^{(2)}/D^{(2)})^2$. This contribution could be significant because $D^{(2)}$ depends on the magnitudes of the cross-Kerr couplings themselves, rather than their differences, cf.~$D^{(1)}$. 

However, since all operations described in this work preserve the total phonon number\footnote{While couplings of the form $(m_{A}m_{B}m_{C}^\dagger+\text{H.c.})$ do not preserve phonon number, the gates we describe only require cyclic evolutions under such couplings, which do preserve total phonon number.}, the detuning $D^{(2)}$ can potentially be reduced by subtracting off the average cross-Kerr shift. Let $\bar{\chi}$ denote the average cross-Kerr coupling across all populated modes, and let $n_{tot}$ denote the total number of phonons in the system. If the drive frequency is chosen to be $\omega_A + \omega_B - \omega_C - \bar{\chi}(n_{tot}-2)$, the detuning becomes\footnote{Because the coupling acts nontrivially only on initial state $\ket{11}_{AB}$, the total number of phonons in modes $j\not\in \{A,B,C\}$ can be assumed to be $n_{tot}-2$.}
\begin{align}
    D'^{(2)}= D^{(2)} +  \bar{\chi}(n_{tot}-2) = \sum_{j}\left[
    \left(\chi_{Cj}-\bar{\chi}\right)
    -\left(\chi_{Aj}-\bar{\chi}\right)
    -\left(\chi_{Bj}-\bar{\chi}\right)
    \right]n_{j}.
\end{align}
$D'^{(2)}$ depends only on deviations in cross-Kerr couplings from the average; when these deviations are small, the infidelity is suppressed. We estimate that the detuning $D'^{(2)}/2\pi$ lies in the range $\approx 100\text{Hz} - 1$kHz, so the associated infidelity is $(g_v^{(2)}/D'^{(2)})^2 \approx 10^{-4} - 10^{-2}$. This infidelity can be even further suppressed using  mode-dependent compensations (e.g.~by leveraging the knowledge that, say, $\lambda_{A}>\lambda_{B}>\lambda_{C}$). We thus conclude that, with appropriate compensations to the drive frequency, the infidelity associated with self- and cross-Kerr interactions can be made negligible relative to the infidelity associated with decoherence and spectral crowding.

In summary, when engineering two-mode couplings, the drive frequencies should be chosen to satisfy the resonance condition
\begin{equation}
    \omega_2 - \omega_1 = \omega_B - \omega_A + S_B - S_A.
\end{equation}
This choice fully compensates the effects of AC Stark Shifts. While the effects of self- and cross-Kerr interactions cannot generally be compensated, the resulting detuning is only a function of small cross-Kerr differences.
Similarly, when engineering three-mode couplings, the drive frequency should be chosen to satisfy the resonance condition
\begin{equation}
    \omega_1 = \omega_A + \omega_B - \omega_C + S_A + S_B - S_C - \chi_{AB} - \bar{\chi} (n_{tot}-2).
    \label{eq:res_cond}
\end{equation}
This choice fully compensates the effects of AC Stark shifts, as well as cross- and self-Kerr terms between the modes $A$, $B$, and $C$. Cross-Kerr interactions with other modes generally contribute to the infidelity. However, the associated detuning from the resonance condition~(\ref{eq:res_cond}) is only a function of small deviations in cross-Kerr, and the infidelity can be further suppressed by using appropriate mode-dependent compensations.
\vspace{2em}

\section{Universality }
In the main text, we focus on the beamsplitter, \texttt{SWAP}, and \texttt{CZ} operations, as these are the only operations we require to implement a QRAM. For completeness, we now show that engineered two- and three-mode interactions can be used to implement a universal gate set. We refer the reader to Refs.~\cite{langford2011,niu2018,niu2018a} for details.

Consider the encoding of a qubit into two phononic modes, $A$ and $B$, as
\begin{equation}
    \ket{\bar{0}} = \ket{1}_A\ket{0}_B,\quad \ket{\bar{1}} = \ket{0}_A\ket{1}_B,
    \label{eq:dual-rail}
\end{equation}
where $\ket{\bar 0, \bar 1}$ denote the qubit states, and $\ket{0,1}_{A,B}$ denote phonon Fock states. Note that this encoding differs from the single-mode encoding discussed in the main text; this dual-rail encoding---commonly considered in optical quantum computing~\cite{knill2001}---is the simplest that allows for universal computation in our scheme. With this encoding, engineered two-mode interactions of the form
\begin{equation}
    H^{(1)} = g_v^{(1)} m_A^\dagger m_B + \text{H.c.}
\end{equation}
can be used to implement arbitrary single-qubit rotations: for $g_{v}^{(1)}$ real (imaginary), $H^{(1)}$ generates rotations around the X (Y) axis of the Bloch sphere. The phase of $g_{v}^{(1)}$ is controllable via the phases of the drives used to engineer the interaction. 

Given two qubits, respectively encoded in modes $\{A_1,B_1\}$ and $\{A_2,B_2\}$ with the encoding~(\ref{eq:dual-rail}), and an ancillary mode $C$ initialized in $\ket{0}_C$, we can implement a \texttt{CZ} gate using the engineered three-mode interaction,
\begin{equation}
    H^{(2)}=g_{v}^{(2)}m_{B_{1}}m_{B_{2}}m_{C}^{\dagger}+\text{H.c.}
\end{equation}
Evolution under $H^{(2)}$ for a time $\pi/g_v^{(2)}$ imparts a -1 phase to the $\ket{\bar 1 \bar 1}$ state, while leaving all other initial states unaffected, thereby realizing the CZ gate. Arbitrary single-qubit rotations and CZ are universal. Note that these engineered two- and three-mode interactions can similarly be used to implement a universal set of logical gates for certain bosonic quantum error-correcting codes~\cite{niu2018a}.


\section{Parallelism and limitations }
One appealing property of virtual gates is that they can be performed in parallel. Parallelism, together with the potential for a high degree of connectivity, makes the proposed architecture highly efficient. In this section, we discuss how virtual gates can be performed in parallel and consider practical limitations on the degree of parallelization possible. 

We first provide a simple example to illustrate how gates may be performed in parallel. Say one wants to implement a \texttt{SWAP} gate between modes $A$ and $B$ in parallel with a \texttt{SWAP} gate between modes $C$ and $D$. A collection of drives must be applied so that the requisite couplings, $(m_{A}^{\dagger}m_{B}+\text{H.c.})$ and $(m_{C}^{\dagger}m_{D}+\text{H.c.})$ are brought on resonance simultaneously. Assuming $|\omega_{B}-\omega_{A}|\neq|\omega_{D}-\omega_{C}|$, this requires a minimum of three distinct drive tones. For example, one can apply three drives whose frequencies satisfy the resonance conditions
\begin{align}
    \omega_{2}-\omega_{1}	&=\omega_{B}-\omega_{A}, \\
    \omega_{3}-\omega_{2}	&=\omega_{D}-\omega_{C},
\end{align}
where for simplicity we have the neglected compensations for Stark shifts and cross-Kerr interactions (see Sec.~\ref{frequency_shifts}).
Importantly, in order to ensure that only the intended interactions are brought on resonance, we must also have $|\omega_{3}-\omega_{1}|\not \approx |\omega_{j}-\omega_{k}|$ for any modes $j,k$ which are used to store quantum information. Provided that this condition is satisfied, the \texttt{SWAP} gates are implemented in parallel without disturbing other modes, and the operation in parallel is not fundamentally different from serial operation. 

The restriction $|\omega_{3}-\omega_{1}|\not \approx |\omega_{j}-\omega_{k}|$ illustrates an important general point: when applying multiple drives to perform gates in parallel, all combinations of the drive frequencies must be carefully checked in order to verify that only the desired couplings are brought on resonance. More specifically, for $d$ drives applied simultaneously, $d(d-1)/2$ pairwise combinations must be checked to ensure that only the desired two-mode couplings are resonant. Furthermore, each of the $d$ frequencies must also be individually checked to ensure that only the desired three-mode couplings are resonant. As the number of drives grows, it is likely to become increasingly challenging to avoid accidentally bringing some unwanted coupling on resonance. The need to avoid unwanted resonant couplings can thus limit on the degree of parallelization possible. This limit generally depends on the specific structure of the nonuniformity, the desired connectivity, and the total number of modes. 

An additional practical limit may come from the need to avoid drive-induced heating of the transmon. As discussed in Ref.~\cite{zhang2019}, 
the applied drives can facilitate an effective ``heating'' of the transmon via its natural dissipation and dephasing processes, and this heating can limit the operation fidelity.  With two drive tones, the rate of this drive-induced heating is~\cite{zhang2019}\footnote{There is a factor of 2 missing in Eq.~(48) in Ref.~\cite{zhang2019}; expression~(\ref{eq:drive_heat}) is the correct expression. }
\begin{equation}
    \label{eq:drive_heat}
    \gamma\left[ 
    \left|\frac{\alpha \xi_1^2}{2\delta_1+\alpha} \right|^2 +
    \left|\frac{\alpha \xi_2^2}{2\delta_2+\alpha} \right|^2 +
    \left|\frac{2\alpha \xi_1\xi_2}{\delta_1+\delta_2+\alpha} \right|^2
    \right] + 
    2\gamma_\phi^{\text{(hf)}}\left[ |\xi_1|^2 + |\xi_2|^2 \right]
\end{equation}
to leading order in $\xi_{1,2}$. Here, $\gamma$ is the transmon loss rate, and $\gamma_\phi^{\text{(hf)}}$ is the pure dephasing rate of the transmon due to noise at high frequencies\footnote{While a typical Ramsey experiment probes the noise spectrum near zero frequency, it is the spectral content at high frequencies $\sim \delta_{1,2}$ that is relevant to drive-induced heating via dephasing. }, which was measured to satisfy $\gamma_\phi^{\text{(hf)}} \ll \gamma$ in Ref.~\cite{zhang2019}.
For the weak drives considered in this work, the infidelity associated with this additional heating is negligible relative to that from the inverse Purcell effect. As more drives are added, however, this effective heating rate will increase and could limit operation fidelity.

\end{document}